\documentclass[%
 reprint,
 amsmath,amssymb,
 aps,
 prd,
 showkeys,
]{revtex4-2}
\usepackage{graphicx}
\usepackage{dcolumn}
\usepackage{bm}
\usepackage{color}

\newcommand{\be}{\begin{eqnarray}}
\newcommand{\ee}{\end{eqnarray}}
\newcommand{\bea}{\begin{eqnarray}}
\newcommand{\eea}{\end{eqnarray}}

\begin{document}

\preprint{APS/123-QED}

\title{Polarization Signatures of Rotating Black Holes in Perfect Fluid Dark Matter Spacetimes}

\author{Yu-Xiang Huang$^{1}$, Sen Guo$^{2}$, En-Wei Liang$^{1*}$, Kuan Liu$^{1}$, Kai Lin$^{3}$}

\affiliation{
$^1$Guangxi Key Laboratory for Relativistic Astrophysics, School of Physical Science and Technology, Guangxi University, Nanning 530004, People's Republic of China\\
$^2$College of Physics and Electronic Engineering, Chongqing Normal University, Chongqing 401331, People's Republic of China\\
$^3$Universidade Federal de Campina Grande, Campina Grande, PB 58429-900, Brasil}

\email{$^1$ yxhuangphys@126.com \\ $^2$ sguophys@126.com \\ $^1$ lew@gxu.edu.cn (Corresponding author) \\ $^1$ liuk@st.gxu.edu.cn \\ $^3$ kailin@if.usp.br }

\date{\today}
\begin{abstract}
Motivated by horizon-scale polarization observations of M87$^{*}$, we investigate polarized emission from rotating black holes (BHs) immersed in a perfect fluid dark matter (PFDM) background. Fully relativistic ray tracing 
is utilized to examine how the polarization structure jointly depends on magnetic-field geometry, higher-order imaging, and the PFDM intensity parameter $k$. We show that, for a given $k$ value, magnetic configurations with a polar component naturally produce a continuous spiral pattern in the electric vector position angle (EVPA), while the large-scale EVPA morphology remains primarily controlled by the magnetic-field topology. Variations in $k$ affect light propagation and polarization transport near the horizon, leading to corresponding changes in EVPA deflection and polarized intensity around the photon ring. A decomposition of the total polarized image further reveals that higher-order images provide localized yet non-negligible corrections near the ring. Compared with the polarization characteristics of M87$^{*}$ inferred by the EHT, image-domain quantities such as $|m|_{\rm net}$ and $\arg\beta_2$ indicate that PFDM induces systematic shifts relative to the pure Kerr BH case, lowering the net linear polarization fraction and modifying the large-scale polarization phase. Overall, PFDM acts as an additional strong-gravity ingredient that systematically reorganizes the horizon-scale polarization structure, suggesting that surrounding dark matter can leave observable imprints on black hole polarization signatures.

\end{abstract}

\maketitle

\section{Introduction}
\label{sec:intro}
\par
The Event Horizon Telescope (EHT) has ushered in a new era of black hole (BH) physics with the landmark imaging of the supermassive BHs Messier 87$^*$ (M87$^*$) and Sagittarius A$^*$ (Sgr A$^*$)~\cite{EventHorizonTelescope:2019dse,EventHorizonTelescope:2022wkp}. 
Beyond the characteristic bright, ring-like emission produced by strong gravitational lensing, the  horizon-scale polarized images by ETH in 2021 and 2024 have opened a new observational window on the magnetic field structure in the immediate vicinity of the event horizon~\cite{EventHorizonTelescope:2021bee,EventHorizonTelescope:2021srq,EventHorizonTelescope:2024hpu,EventHorizonTelescope:2024rju}. 
Multi-epoch observations spanning 2017–2021 further indicate that, while the ring diameter remains remarkably stable, the linear polarization fraction and the Electric Vector Position Angle (EVPA) exhibit significant variability and even helicity reversals~\cite{EventHorizonTelescope:2025vum}. These polarization signatures indicate that horizon-scale magnetic fields possess nontrivial spatial structure, making polarization a uniquely sensitive probe of near-horizon plasma conditions and magnetic geometry.

These observations give insights into the polarized light propagation in curved spacetime. 
The covariant transport of polarization vectors along null geodesics, formulated by Beloborodov~\cite{beloborodov2002gravitational}, provides the foundation for modern relativistic polarization modeling. Building on this framework, Narayan \textit{et al.} recently constructed models that successfully reproduce polarization features observed in M87$^*$ using equatorial synchrotron emission in Schwarzschild geometry~\cite{narayan2021polarized}. Gelles \textit{et al.} extended this to rotating BH~\cite{gelles2021polarized}. More recently, polarization has also been explored as a sensitive probe of departures from vacuum Kerr geometry, including modified gravity theories and the presence of exotic matter distributions~\cite{Qin:2022kaf,Zhang:2022klr,Delijski:2022jjj,zhang2024polarized,huang2024coport}. Furthermore, general relativistic magnetohydrodynamic (GRMHD) simulations have shown that the large-scale geometry of the magnetic field plays an equally crucial role. In particular, organized polar or poloidal magnetic field components near the event horizon are essential for producing coherent EVPA patterns, polarization asymmetries, and lensing-induced structures seen in horizon-scale images~\cite{ricarte2021black,nathanail2022magnetic}. The observed EVPA coherence and spiral-like polarization patterns therefore point to the central role of polar magnetic field components in organizing the horizon-scale polarization structure. 

While most BH polarization studies are based on vacuum Kerr spacetime and modified-gravity BHs, the influence of dark matter remains largely unexplored. In the standard cosmological model, dark matter constitutes approximately 27\% of the universe's energy content. Although it has not been directly detected, its gravitational influence is well established~\cite{Bahcall:1991qs,Zaritsky:1996ch,Mateo:1998wg,Koopmans:2002qh}.Recent dynamical studies of M87$^*$ reveal that dark matter plays a critical role in shaping its mass distribution. Multi-tracer analyses have confirmed a dark matter fraction exceeding 80\% at large radii~\cite{murphy2011galaxy}, and have revealed a dominant dark halo with a core-like structure inferred from both stellar and globular cluster dynamics~\cite{oldham2016galaxy}. These findings strongly suggest that dark matter may significantly affect the inner dynamics and observed emission of the central supermassive BH. A useful starting point for incorporating a surrounding fluid into BH spacetimes is the static spherically symmetric solution obtained by Kiselev~\cite{Kiselev:2002dx}, originally derived for quintessence and therefore formulated in a dark-energy context. Related Kiselev-type fluid solutions were later applied to dark-matter phenomenology, including galactic rotation curves~\cite{kiselev2003quintessential} and galactic dark-matter configurations with a central supermassive BH~\cite{Li:2012zx}, and this effective-fluid description was subsequently generalized to rotating BHs~\cite{Xu:2017bpz}. This rotating PFDM solution provides a useful theoretical basis for investigating how dark matter might influence the polarized emission around BHs. 

Motivated by these considerations, we present a systematic study of polarized radiative transfer around rotating BHs in PFDM spacetimes, focusing on the horizon-scale polarization structure of M87$^{*}$. Our goal is to disentangle the effects of magnetic-field topology and PFDM-modified spacetime geometry on the observed polarization pattern. To quantify the resulting Kerr--PFDM differences beyond visual morphology alone, we introduce image-domain diagnostics, in particular $|m|_{\rm net}$ and $\arg\beta_2$, within an observationally relevant framework.

The structure of this paper is as follows. In Sec.~\ref{sec:2}, we present the theoretical framework, including the rotating PFDM spacetime, the equations of photon motion, and the covariant transport of polarization. In Sec.~\ref{sec:3}, we investigate the polarization features produced by different magnetic-field configurations, analyze how the PFDM parameter modifies the EVPA structure and polarized intensity, and assess the role of higher-order images in shaping the total polarization map. In Sec.~\ref{sec:4}, we place these results in a quantitative and observational context by connecting the PFDM parameter to a representative physical density scale for M87$^{*}$, introducing the image-domain observables $|m|_{\rm net}$ and $\arg\beta_2$, and examining how the corresponding Kerr--PFDM differences are realized in comparison with the observed polarization organization of M87$^{*}$. Finally, in Sec.~\ref{sec:5}, we summarize our main results and discuss their implications for horizon-scale polarimetry and for probing dark matter effects in the strong-gravity regime.

\section{Theoretical Framework}
\label{sec:2}
\par
To incorporate the effects of PFDM on a rotating BH, we adopt a minimally–coupled prescription in which the PFDM is described by an anisotropic perfect–fluid stress tensor and sources the Einstein equations. The total action reads \cite{kiselev2003quintessential,Li:2012zx}:
\begin{align}
	\label{eq:1}
	R_{\mu\nu} - \frac{1}{2}g_{\mu\nu} R = 8\pi (T_{\mu\nu}^{M} + T_{\mu\nu}^{DM}) \equiv 8\pi T_{\mu\nu},
\end{align}
where the interaction between baryonic ($T_{\mu\nu}^{M}$) and dark matter ($T_{\mu\nu}^{DM}$) components is absorbed into the total stress-energy tensor. A stationary, axisymmetric rotating BH solution embedded in PFDM can be written in Boyer–Lindquist coordinates \((t,r,\theta,\phi)\) as \cite{Xu:2017bpz}:
\begin{align}
	\label{eq:2}
	{\rm d}s^{2}= &-\Bigg(1-\frac{2m(r)r}{\Sigma(r)}\Bigg){\rm d}t^{2} + \frac{\Sigma(r)}{\Delta(r)}{\rm d}r^{2} \nonumber\\
	&+ \Sigma(r){\rm d}\theta^{2}- \frac{4m(r)ra\sin^{2}\theta}{\Sigma(r)}{\rm d}t{\rm d}\phi \nonumber\\
	&+ \Bigg[\sin^{2}\theta(r^{2}+a^{2})+\frac{2m(r)r a^{2}\sin^{4}\theta}{\Sigma(r)}\Bigg] {\rm d}\phi^{2},
\end{align}
with
\begin{align}
	\label{eq:3}
	&\Sigma(r) =r^{2} + a^{2}\cos^{2}\theta,\\
	\label{eq:4}
	&\Delta(r) = r^{2} + a^{2} - 2m(r)r, \\
	\label{eq:5}
	& m(r)=M -\frac{k}{2} \ln \frac{r}{\mid k \mid},
\end{align}
where $M$ and $a$ are the mass and spin parameter of BHs and $k$ parametrizes the PFDM intensity. In an orthonormal basis, the stress-energy tensor of PFDM is diagonal \cite{Xu:2017bpz},
\begin{align}
	\label{eq:6}
	T^{\hat{\mu}}{}_{\hat{\nu}} = \mathrm{diag}(-\rho,\, p_r,\, p_\theta,\, p_\phi),
\end{align}
with components
\begin{align}
	\label{eq:7}
	\rho & = \frac{k r}{8\pi (r^2 + a^2 \cos^2 \theta)^2}, \\
	p_r & = -\rho, \\
	p_\theta & = p_\phi = -p_r - \frac{k}{16\pi r (r^2 + a^2 \cos^2 \theta)}.
\end{align}
This proportionality shows that $k$ directly sets the overall scale of the effective PFDM energy density: a larger $k$ corresponds to a denser PFDM distribution around the BH, while $k\to 0$ recovers the vacuum Kerr spacetime. In geometrized units ($G=c=1$), $k$ has dimension of length (or equivalently, mass), and therefore serves as a well-defined intensity parameter for PFDM in both static and rotating BH configurations. A representative conversion of the adopted $k$ range into an effective physical density scale, for the case of M87$^{*}$, will be discussed later in Sec.~\ref{sec:4}.

\subsection{Equations of Photon Motion}
\label{sec:2-1}
\par
The morphology of the BH shadow and the structure of the surrounding accretion flow are predominantly governed by the spacetime geometry, which is in turn determined by the BH parameters $(M,a)$ and the PFDM parameter $k$. To study photon trajectories in such a rotating PFDM spacetime, we employ the Hamilton--Jacobi formalism. Owing to the separability of the Hamilton--Jacobi equation in axisymmetric geometries, the geodesic equations can be derived using the conserved quantities associated with stationarity, axisymmetry, and the Carter constant. The Hamilton--Jacobi equation takes the form \cite{carter1968global,chandrasekhar1998mathematical}
\begin{align}
\label{eq:8}
2\frac{{\partial S}}{{\mathrm{d}\sigma}}  = -g^{\mu \nu}  \frac{\partial S}{\partial x^{\mu}}\frac{\partial S}{\partial x^{\nu}},
\end{align}
where $\sigma$ is the affine parameter and $S$ denotes the Jacobi action, written as
\begin{align}
\label{eq:9}
S= S_{r}(r)+S_{\theta}(\theta) + L \phi-E t.
\end{align}

The geodesic motion is then characterized by three independent constants of motion:
the energy $E=-p_{t}$, the axial angular momentum $L=p_{\phi}$, and the Carter constant $\mathcal{C}$, which together determine the full four-momentum $p^{\mu}$. The photon momentum components follow \cite{chandrasekhar1998mathematical}
\begin{align}
	\label{eq:10}
	&\frac{\Sigma}{{\mathrm{d}\sigma}}p^{r} = \pm_{r} \sqrt{\mathcal{A}(r)}, \\
	\label{eq:11}
	&\frac{\Sigma}{{\mathrm{d}\sigma}}p^{\theta} = \pm_{\theta} \sqrt{\mathcal{B}(\theta)}, \\
	\label{eq:12}
	&\frac{\Sigma}{{\mathrm{d}\sigma}}p^{\phi} =
	\frac{a}{\Delta}(E r^{2}+E a^{2}-L a)+\frac{L}{\sin^{2}\theta}-E a, \\
	\label{eq:13}
	&\frac{\Sigma}{{\mathrm{d}\sigma}}p^{t} =
	\frac{a^{2}+r^{2}}{\Delta(r)}(E a^{2}+E r^{2}-L a)+L a-E a^{2}\sin^{2}\theta,
\end{align}
where $\pm_r$ and $\pm_\theta$ indicate the propagation directions in the radial and polar coordinates, respectively. The radial and polar effective potentials are
\begin{align}
	\label{eq:14}
	&\mathcal{A}(r) = -\Delta(r) [(E a-L)^{2}+\mathcal{C}]
	+(E a^{2}+E r^{2}- L a)^{2}, \\
	\label{eq:15}
	&\mathcal{B}(\theta) =\mathcal{C}- L^{2}\cot^{2}\theta + E^{2}a^{2}\cos^{2}\theta.
\end{align}

To better expose their geometrical interpretation, we introduce the two standard impact parameters \cite{carter1968global}
$$
\label{eq:16}
\xi = \frac{L}{E}, \qquad \eta = \frac{\mathcal{C}}{E^{2}},
$$
which uniquely describe photon trajectories as seen by an asymptotic observer. Substituting these into Eqs.~(\ref{eq:14})--(\ref{eq:15}) simplifies the analysis of circular photon orbits. For photons located at $r=r_{p}$, the critical values $(\xi,\eta)$ corresponding to unstable circular photon orbits satisfy $\mathcal{A}=0$ and $\partial \mathcal{A}/\partial r=0$, yielding
\begin{align}
	\label{eq:17}
	&\xi=\frac{(2f(r)+r f'(r))(a^{2}+r^{2})-4a^{2}-4f(r)r^{2}}{2af(r)+arf'(r)}, \\
	\label{eq:18}
	&\eta=\frac{r^{3}[8a^{2}f'(r)-(-2f(r)+rf'(r))^{2}r]}
	{(2f(r)+rf'(r))^{2}a^{2}},
\end{align}
where $f(r)=1-\frac{2M}{r} -\frac{k}{r} \ln \frac{r}{\mid k \mid}$ and $f'(r)$ is its radial derivative. For an observer at asymptotic infinity, the celestial coordinates $(\alpha,\beta)$ associated with a photon arriving from direction $(\xi,\eta)$ are \cite{chandrasekhar1998mathematical}
\begin{align}
	\label{eq:19}
	&\alpha=-\frac{\xi}{\sin \theta}, \\
	\label{eq:20}
	&\beta=\pm \sqrt{a^{2}\cos^{2}\theta+\eta-\xi^{2}\cot^{2}\theta}.
\end{align}
These relations map the constants of motion directly to observable positions on the image plane and thus determine the silhouette of the BH shadow.

\subsection{Theory of Polarization}
\label{sec:2-2}
\par
In analyzing synchrotron polarization in BH accretion flows, it is essential to introduce a local reference frame in which the emission and polarization properties are defined. This frame is not the frame of a distant observer, but the local orthonormal tetrad comoving with the plasma fluid. The fluid four-velocity $u^{\mu}$ specifies this comoving frame, where both the photon wave vector $\hat{k}$ and the magnetic field $\hat{B}$ appear as spatial three-vectors orthogonal to $u^{\mu}$. Within this local fluid rest frame, the instantaneous polarization direction is determined by
\begin{align}
\label{eq:21}
\hat{f}^{\mu}=\hat{\epsilon}^{\mu \nu \rho}\hat{k}_{\nu}\hat{B}_{\rho},
\end{align}
where $\hat{\epsilon}^{\mu \nu \rho}$ is the Levi–Civita tensor associated with the comoving orthonormal frame. It encodes the antisymmetric cross-product structure that relates the photon propagation direction to the magnetic field in three-dimensional space, allowing the intrinsic polarization direction to be constructed locally. This provides an accurate description of the emission physics and captures how the polarization angle is shaped by the magnetized plasma near the BH.

However, such a three-dimensional construction is inherently tied to the local fluid rest frame and cannot be directly propagated through curved spacetime. For GR radiative transfer, the polarization must be expressed in a fully covariant four-dimensional form so that it can be consistently evolved along null geodesics. The transition from the intrinsic 3D definition to its 4D spacetime counterpart is achieved through the relation
$\hat{\epsilon}^{\mu \nu \rho} = \epsilon^{\mu \nu \rho \varrho} u_{\varrho}$,
where $\epsilon^{\mu \nu \rho \varrho}$ is the four-dimensional Levi–Civita tensor. This identification promotes the intrinsic 3D cross-product structure into a covariant expression that remains valid in arbitrary coordinates and along the entire photon trajectory.

Replacing the spatial vectors by their spacetime counterparts then yields the covariant polarization vector,
\begin{align}
	\label{eq:22}
	f^{\mu}=\epsilon^{\mu \nu \rho \varrho} u_{\nu} k_{\rho} B_{\varrho},
\end{align}
which is orthogonal to both $u^{\mu}$ and $k^{\mu}$ by construction. This four-dimensional expression is essential for consistently embedding the locally defined emission physics into the global spacetime geometry, ensuring that the polarization information is expressed in a form suitable for propagation along null geodesics.

Our analysis ultimately seeks the polarization vector measured by an observer at spatial infinity. Once the intrinsic emission vector $f^{\mu}$ is defined by Eq.~(\ref{eq:22}), its subsequent evolution is governed by the parallel transport equation \cite{carroll2019spacetime}
\begin{equation}
	\label{eq:23}
	\frac{\rm d}{{\rm d}\sigma}\mathcal{F}^{\mu}
	+ \Gamma^{\mu}_{\nu\rho}\frac{{\rm d}x^{\nu}}{{\rm d}\sigma}\mathcal{F}^{\rho}=0,
\end{equation}
where $\Gamma^{\mu}_{\nu\rho}$ are the Christoffel symbols of the background metric and $\mathcal{F}^{\mu}$ is the parallel-transported polarization vector. Imposing $\mathcal{F}^{\mu}(\sigma_0)=f^{\mu}$ at the emission point guarantees a self-consistent tracking of the polarization state from the local plasma frame to the distant observer.

To relate the parallel-transported polarization vector to observable quantities, we next introduce the orthonormal basis associated with a Zero Angular Momentum Observer (ZAMO). The ZAMO frame is particularly suitable for constructing an image-plane polarization map because it represents a locally nonrotating observer whose four-velocity is orthogonal to the spatial hypersurfaces of constant Boyer–Lindquist time. As a result, ZAMOs provide a natural local rest frame free from frame-dragging effects, allowing physical quantities such as intensity and polarization to be projected in a manner that is directly comparable to measurements made by a distant observer. Furthermore, the ZAMO tetrad forms a fully orthonormal screen basis, ensuring that the projection of the transported polarization vector preserves its geometric and normalization properties. In this frame, a fisheye projection is employed to map the incoming geodesics onto the observer’s image plane. The screen is spanned by the orthonormal tetrad vectors \cite{Hu:2020usx}
\begin{align}
	\label{eq:24}
	&e_{(\rm t)}=\varpi \partial_{t} + \lambda \partial_{\phi},\qquad
	e_{(\rm r)}=\frac{1}{\sqrt{g_{\rm rr}}}\partial_{r},\\
	&e_{(\theta)} = -\frac{1}{\sqrt{g_{\theta\theta}}} \partial_{\theta}, \qquad 
	e_{(\phi)} = \frac{1}{\sqrt{g_{\phi\phi}}} \partial_{\phi}.
\end{align}
where
\begin{equation}
	\label{eq:25}
	\varpi = \sqrt{\frac{g_{\rm \phi\phi}}{g^{2}_{\rm t\phi}- g_{\rm tt}g_{\rm \phi\phi}}},~~~\lambda=-\frac{g_{\rm t\phi}}{g_{\rm \phi\phi}}\sqrt{\frac{g_{\rm \phi\phi}}{g^{2}_{\rm t\phi}- g_{\rm tt}g_{\rm \phi\phi}}}.
\end{equation}
To connect the parallel-transported polarization vector with observable quantities, we project $\mathcal{F}^{\mu}$ onto the screen coordinates defined by the orthonormal basis in Eqs.~(\ref{eq:24})--(\ref{eq:25}). This yields the measurable polarization components in the image plane \cite{dexter2016public}:
\begin{align}
	\label{eq:26}
	f_n^{(\alpha)} = \mathcal{F}^{\mu} e_{(\phi) \mu}, \quad 
	f_n^{(\beta)} = \mathcal{F}^{\mu} e_{(\theta) \mu}.
\end{align}
Here, \(f_n^{(\alpha)}\) and \(f_n^{(\beta)}\) represent the polarization vector components along the azimuthal and polar directions of the observer’s screen, respectively. Since our radiative model assumes a geometrically thin emitting disk on the equatorial plane, only photon trajectories intersecting this emitting surface contribute to the observed polarized flux. By summing over all such geodesics and their equatorial crossings, the Stokes parameters \(Q\) and \(U\) follow from their standard definitions \cite{huang2024coport}:
\begin{align}
	\label{eq:27}
	Q &= \sum_{n=1}^{n_{\rm max}} g^3 J_{\rm model} \left[ (f_n^{(\alpha)})^2 - (f_n^{(\beta)})^2 \right], \\
	\label{eq:28}
	U &= \sum_{n=1}^{n_{\rm max}} g^3 J_{\rm model} \left[ 2 f_n^{(\alpha)} f_n^{(\beta)} \right],
\end{align}
where $n$ denotes the number of times the light intersects the equatorial plane, $g$ is the redshift factor, and $J_{\rm model}$ represents the emissivity profile on the equatorial plane. These quantities are given by \cite{Chael:2021rjo}:
\begin{align}
	\label{eq:29}
	g &= \frac{\nu_{\mathrm{o}}}{\nu_{\mathrm{s}}} = \frac{E}{E_{\mathrm{s}}},\\
	\label{eq:30}
	\log[J_{\mathrm{model}}(r)] &= -2 \log(r/r_{\mathrm{H}}) - \frac{1}{2} \left[\log(r/r_{\mathrm{H}})\right]^2, 	
\end{align}
where $r_{\mathrm{H}}$ denotes the horizon radius, $\nu_{\mathrm{o}}$ and $\nu_{\mathrm{s}}$ are the observed and source-frame photon frequencies, $E$ and $E_{\mathrm{s}}$ are the corresponding energies. For light propagation in vacuum, the quantity $I_{\nu}/\nu^{3}$ is conserved along geodesics \cite{Lindquist:1966igj}. Consequently, the observed intensity can be computed by integrating along the light ray trajectory as \cite{Chael:2021rjo}
\begin{equation}
	I_{\nu_{\mathrm{o}}} = \sum\limits_{s = 1}^{N_{\mathrm{max}}} f_s g^3(r_s) J_{\mathrm{model}}(r_s),
\end{equation}
where $f_s$ is a fudge factor adjusting the brightness of higher-order rings \cite{Gralla:2019drh}. For our analysis, we select $f_{\rm s} = 1$ \cite{Hou:2022eev}. Furthermore, using Eq.~\eqref{eq:27} and \eqref{eq:28}, one can readily compute both the total intensity of the linearly polarized light $P$ and EVPA $\chi$, defined respectively as
\begin{align}
\label{eq:31}
P= \sqrt{Q^2 + U^2},\quad \chi=\frac{1}{2}\arctan \frac{U}{Q}.
\end{align}
In our equatorial thin-disk model, the polarized emission receives contributions from both the direct and higher-order images. Accordingly, unless otherwise specified, all polarization maps shown in Sec.~III represent the total image, defined as the sum of these two components. A more detailed discussion of their respective properties and interplay will be given in Sec.~\ref{sec:3-3}.

\section{Polarization Features in PFDM Spacetime}
\label{sec:3}
For modeling the polarization signatures in PFDM spacetimes, we adopt a simplified emission framework that retains the essential kinematic and magnetic ingredients while minimizing unnecessary complications. The emitting plasma is assumed to follow circular Keplerian motion in the equatorial plane, characterized by an angular velocity $\Omega(r)\equiv u^{\phi}/u^{t}$. This quantity reproduces the standard Keplerian frequency in the Newtonian limit and, in curved spacetime, corresponds to the orbital angular velocity as measured by an observer at infinity. The magnetic field is assumed to be comoving with the plasma and purely spatial in the fluid rest frame, i.e., $B^\mu u_\mu=0$. This condition corresponds to the geometric content of the ideal MHD assumption $F^{\mu\nu}u_\nu=0$, but we do not model the dynamical evolution implied by flux freezing. Instead, we prescribe a normalized magnetic-field configuration, $B=(B_r,B_\theta,B_\phi)$, to isolate the impact of field geometry on polarization, and refer to it simply as $B$ throughout the paper. In the PFDM-modified spacetime, this simplified model allows us to highlight how changes in the spacetime structure, combined with a prescribed magnetic field geometry, impact the measurable polarization patterns.

\begin{figure*}[htbp]
	\centering
	\includegraphics[width=5.5cm,height=5.5cm]{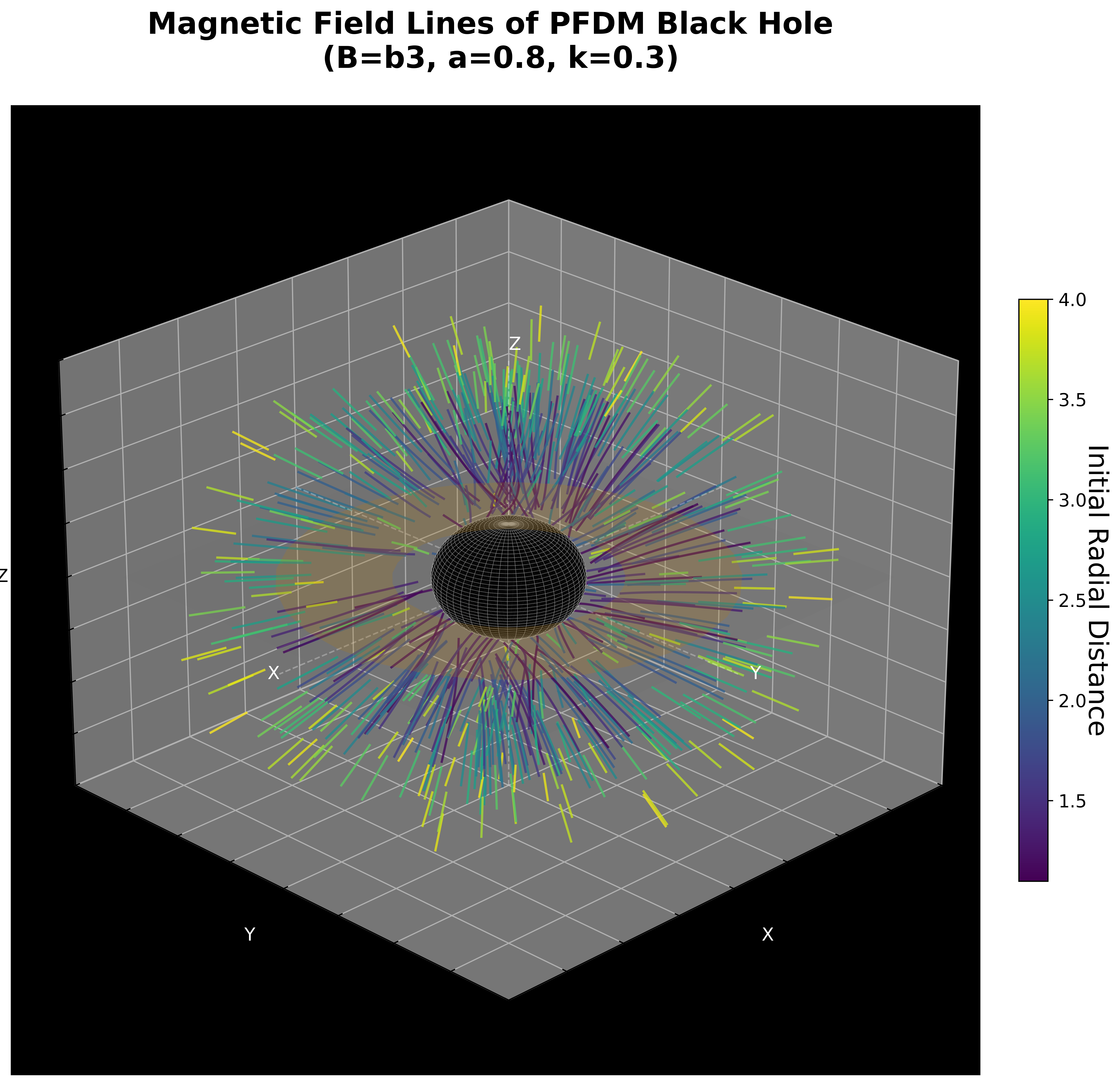}
	\includegraphics[width=5.5cm,height=5.5cm]{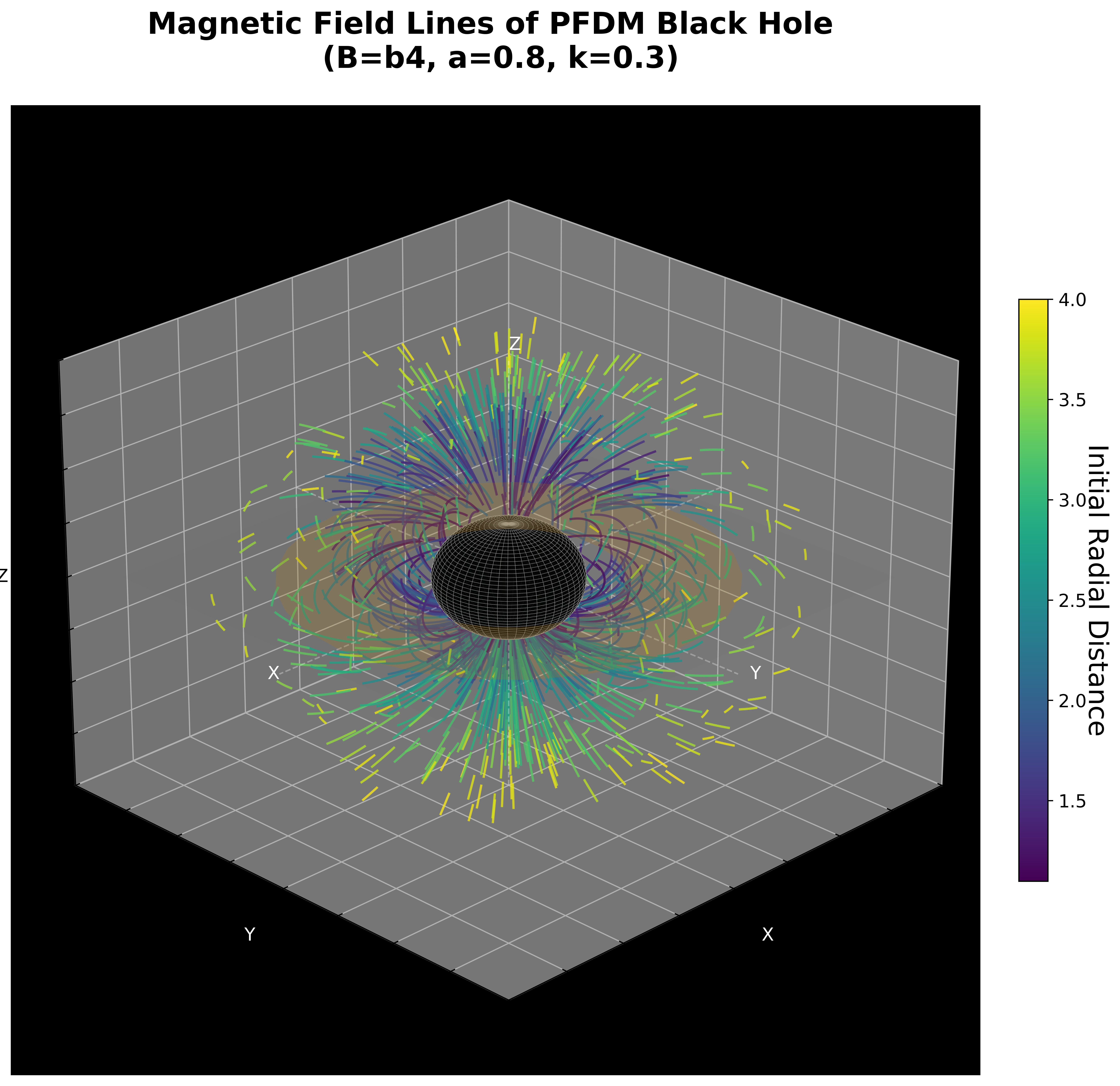}
	\caption{\label{fig:1} Schematic of the two large-scale magnetic-field configurations adopted in this work. Left panel: The third configuration (b3); Right panel: The fourth configuration (b4).}
\end{figure*}

\subsection{Polarization Mapping Under Different Magnetic Fields}
\label{sec:3-1}
A central objective of this study is to examine how different large-scale magnetic-field geometries shape the resulting polarization patterns. Therefore, we investigate four magnetic-field configurations and assess their respective impacts on the polarization morphology. The first configuration, introduced in Ref.~\cite{narayan2021polarized}, includes radial and azimuthal components and is given by $B = (0.87, 0, 0.5)$. Within the fluid frame used in that work, this setup was found to most successfully reproduce the observed polarization morphology of M87$^{*}$. The second configuration is introduced here by replacing the azimuthal component of the EHT model with a polar one, while preserving the overall strength distribution: $B = (0.87, 0.5, 0)$. This variant isolates the influence of a polar magnetic component on the projected polarization structure. The third configuration probes the theoretical limit of a large-scale, ordered poloidal field threading the BH---a structure central to jet-launching mechanisms such as Blandford--Znajek \cite{blandford1977electromagnetic}. To isolate its unique polarization signature, we employ an idealized, magnetically dominated field geometry that follows the initial setup used in the GRMHD simulations of Komissarov \cite{Komissarov:2009dn}. This setup represents a clean, end-member scenario in which a nearly force-free, global poloidal field is anchored to the black hole, allowing us to study how such an ordered field imprints itself on the polarization signal. The field is constructed in Kerr--Schild coordinates, ensuring regularity across the horizon, and its components, as specified in \cite{Komissarov:2009dn} for the initial data, are:
\begin{align}
	\label{eq:32}
	&B^{r} = B_0 \frac{\sin \theta}{\sqrt{-g}}, \\
	&B^{\phi} = \frac{a \Sigma(r) B^{r} + 2 a r B^{r}}{\Sigma(r) \Delta(r) + 2 r^{3} + a^2 r}, \\
	&B^{\theta} = 0,
\end{align}
where $g$ represents the determinant of the metric tensor, and $B_0$ is the initial magnetic field strength. This contravariant representation ($B^\mu$) is standard in the 3+1 formalism of GRMHD and is distinguished from the orthonormal-component representation $B=(B_{r},B_{\theta},B_{\phi})$ used elsewhere in this work for analyzing local field geometry. Complementing this horizon-threading limit, the fourth configuration describes a highly idealized, static dipolar magnetic field anchored within the inner accretion flow and artificially restricted to the region outside the event horizon. This setup serves as a conceptual counterpoint to the global magnetospheric models by representing a scenario where ordered magnetic structure is confined to the disk environment. To construct this test field, we use the vacuum solution for a pure magnetic dipole in flat spacetime, translated into Cartesian coordinates for numerical convenience:
\begin{align}
	\label{eq:33}
	B^x &= \frac{3xz}{(x^2 + y^2 + z^2)^{5/2}}, \nonumber\\
	B^y &= \frac{3yz}{(x^2 + y^2 + z^2)^{5/2}}, \nonumber\\
	B^z &= \frac{3z^2 - (x^2 + y^2 + z^2)}{(x^2 + y^2 + z^2)^{5/2}}.
\end{align}
We emphasize that this is a highly simplified, non-dynamical proxy intended to isolate the polarization effects of a hypothetical, large-scale poloidal field component associated with the accretion disk, distinct from the turbulent, small-scale fields typically generated in full GRMHD simulations. To enforce the condition of zero magnetic flux through the event horizon, the field is simply set to zero for $r \leq r_+$. This prescription provides a clear, albeit idealized, representation of a scenario in which the ordered magnetic structure is effectively confined to the accretion flow and disconnected from the black hole. By comparing this configuration with the horizon-threading case, we explore how polarization signatures respond to different magnetic topologies, thereby providing a framework for interpreting the complex magnetic morphology observed in systems like M87$^{*}$. For clarity, we refer to the four configurations as $b1$, $b2$, $b3$, and $b4$ throughout the rest of this work.

\begin{figure}[htbp]
	\centering
	\includegraphics[width=5.5cm,height=6cm]{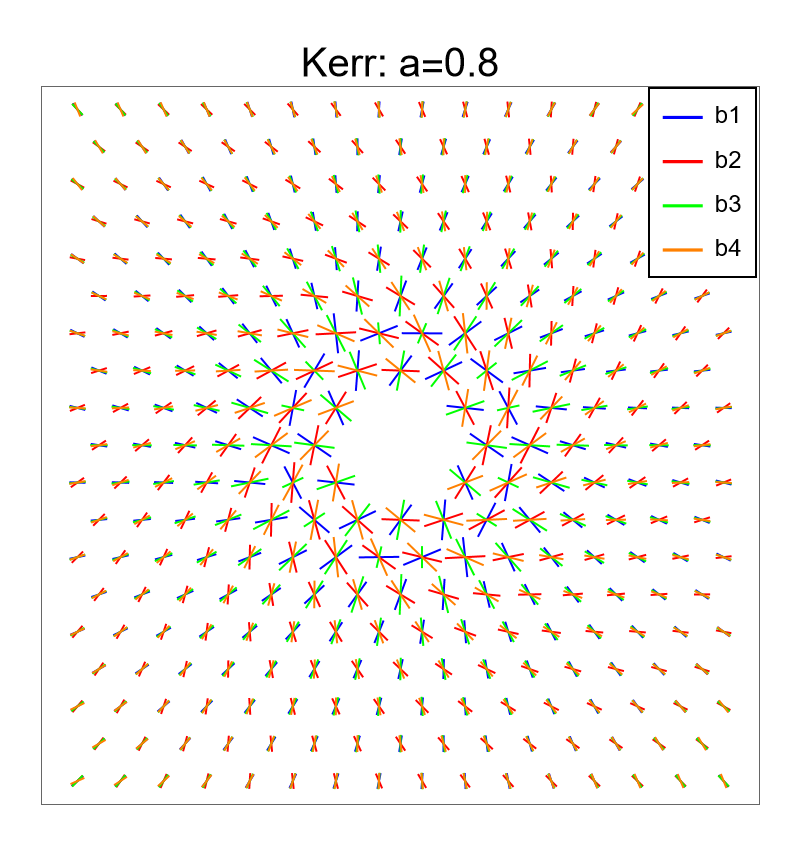}
	\caption{\label{fig:2} Comparison of normalized polarization vectors for four magnetic field configurations ($b_1$, $b_2$, $b_3$ and $b_4$) in a Kerr BH ($a=0.8$). The mass of the BH is set as M = 1.}
\end{figure}

Fig.1 summarizes the global topology of b3 and b4. The left panel illustrates the ordered poloidal field associated with the analytical solution of Ref.~\cite{Komissarov:2009dn}, while the right panel shows the disk-confined dipole with vanishing horizon flux. These schematic diagrams provide visual context for the distinct magnetic regimes represented by the two models.

To systematically compare the polarization effects induced by different magnetic-field configurations, we normalize the computed polarization vectors. This normalization eliminates variations in absolute intensity, allowing a direct comparison of the spatial distribution of polarized intensity and the corresponding evolution of EVPA among different models. Fig. 2 presents the normalized polarization patterns for the four magnetic configurations (b1–b4) in a Kerr spacetime with spin $a = 0.8$, viewed face-on.

A clear trend emerges when examining how the EVPA responds to the underlying magnetic topology. In configuration b1, the EHT-adopted model, the mixed $(B_r,B_\phi)$ structure generates significant EVPA twisting near the BH. However, the azimuthal component also introduces relatively rapid changes in the projected magnetic direction, making the EVPA pattern appear less orderly across the image. By contrast, configuration b2 exhibits a more regular EVPA field. Replacing the azimuthal component with a polar one introduces a large-scale poloidal structure whose projection onto the photon wave vector changes gradually across the image, thereby producing a more organized spiral morphology. The physical origin of this difference follows directly from the polarization relation in Eq.~(\ref{eq:22}), which shows that the EVPA is determined by the magnetic field projected onto the screen plane orthogonal to the photon wave vector. A coherent poloidal component generates a correspondingly coherent projected field, so that the EVPA orientation changes in a less fragmented way from one region of the image to another. Importantly, this interpretation is qualitatively consistent with horizon-scale polarimetry of M87$^{*}$. The EHT Collaboration has demonstrated that the observed linear polarization pattern around M87$^{*}$ exhibits a remarkably coherent azimuthal morphology, which semi-analytic models reproduce only when a substantial poloidal (vertical or radial) field component is included \cite{EventHorizonTelescope:2021srq}. Furthermore, the conversion of an initially vertical magnetic field into an extended spiral pattern, driven by gravitational light bending, relativistic aberration, and frame dragging (see Fig.~3 of \cite{EventHorizonTelescope:2021srq}), mirrors the geometric stabilization and coherent parallel-transport behavior described above. In this sense, the comparison between b1 and b2 does not imply that one magnetic topology is intrinsically superior to the other; rather, it shows that these two idealized configurations leave different EVPA signatures in the image plane. Within the present framework, a substantial poloidal component more readily produces an EVPA pattern with a clearer azimuthal organization, whereas a stronger azimuthal field component tends to introduce more local directional changes.

Configuration $b3$ also contains a substantial poloidal magnetic component, but its resulting polarization morphology differs markedly from those of $b2$ and $b4$. In this case, the analytic magnetospheric field exhibits a highly regular and smoothly varying geometry, such that the projected magnetic orientation sampled by photon trajectories changes only weakly across the emitting region. As a consequence, the resulting EVPA pattern remains relatively compact and shows only limited bending near the BH. Configuration $b4$, by contrast, features a dipolar magnetic field confined to the inner accretion flow, with a strong radial variation of field orientation across the emitting region. Photons originating from different radii therefore encounter systematically different poloidal directions before undergoing relativistic transport. This spatial variation produces a larger change in the projected polarization direction across the image, leading to a more pronounced spiral pattern than in configuration $b3$.

The comparison between $b3$ and $b4$ demonstrates that the presence of a poloidal magnetic component alone is not sufficient to determine the polarization morphology. Rather, what matters is how the poloidal field is distributed within the emitting region and how its orientation varies along photon paths. Configurations with stronger spatial variation in the poloidal field tend to generate a more pronounced spiral EVPA pattern, whereas more uniform poloidal geometries produce a more localized polarization response.
\par


\subsection{Impact of PFDM on Polarization Patterns}
\label{sec:3-2}
\par
\begin{figure*}[htbp]
	\centering
	\includegraphics[width=5.5cm,height=6cm]{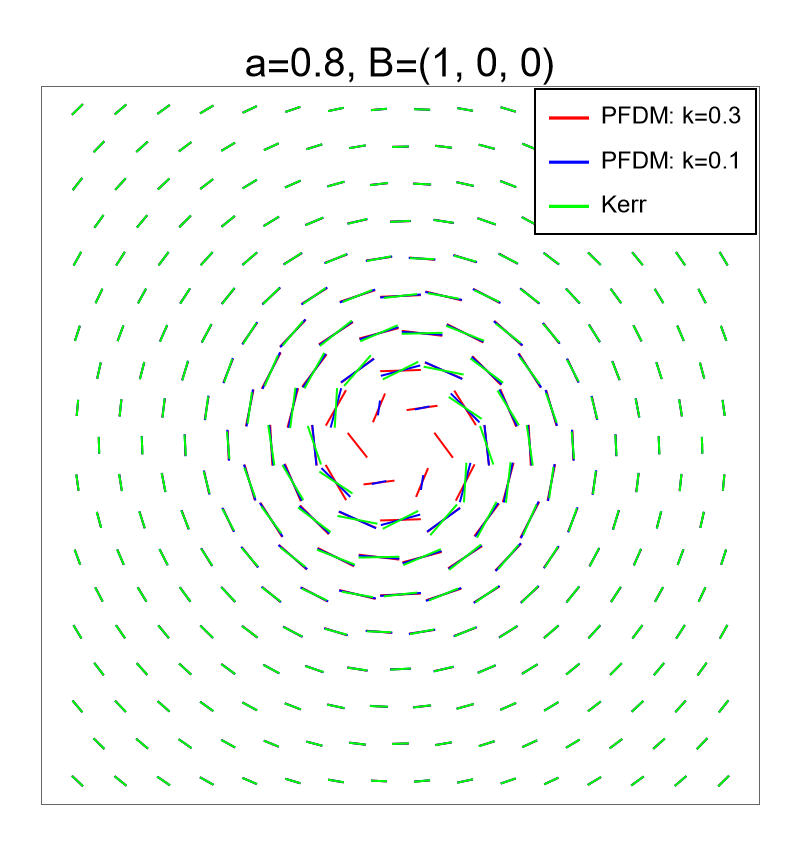}
	\includegraphics[width=5.5cm,height=4cm]{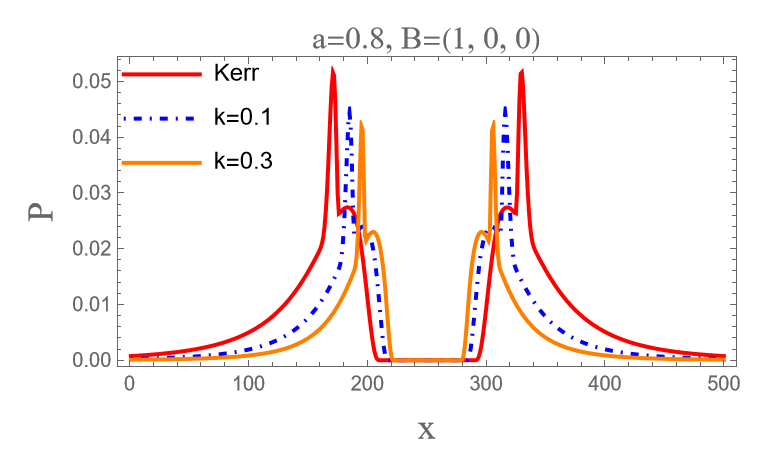}
	\caption{\label{fig:3} Effect of PFDM on polarization with a purely radial magnetic field $B_r = 1$. Left panel: EVPA distribution for Kerr and PFDM BHs ($k = 0$, 0.1, 0.3). Right panel: corresponding polarization intensity $P$. The mass of the BH is set as M = 1.}
\end{figure*}
Having examined how magnetic-field geometry regulates the overall structure and coherence of the polarization map, we now turn to the distinct role played by the PFDM intensity $k$. To isolate the influence of PFDM from magnetic effects, we consider a controlled setup in which only the radial magnetic component $B_r$ is retained. This minimal configuration suppresses azimuthal or polar field-induced structure, ensuring that any changes in the polarization pattern can be attributed primarily to variations in $k$.

We compare a series of BH models with identical spin but different PFDM intensities, including the Kerr case $k = 0$. The left panel of Fig.~3 shows how the EVPA distribution responds to increasing $k$, while the right panel illustrates the one-dimensional variation of the polarization intensity $P$ extracted along the central vertical image cut, denoted as the $x$-axis (with $y=0$ corresponding to the horizontal midline of the image). As $k$ increases from 0.1 to 0.3, the EVPA exhibits a systematic decrease across the image, accompanied by an overall reduction in $P$. Moreover, the PFDM-induced reduction in the horizon radius enhances light deflection near the BH, amplifying strong-gravity effects and thereby making the polarization morphology more sensitive to both spacetime curvature and magnetic-field geometry.

\par
We now analyze the full polarization signatures of the four magnetic-field models in the presence of PFDM. As a representative example, we begin with configuration $b2$, whose strong poloidal component makes it particularly sensitive to spacetime curvature and therefore well suited for isolating the effects of PFDM and BH spin. Fig.~4 shows the polarization vectors superimposed on the total emission for several parameter choices, while Fig.~5 displays the corresponding polarized-intensity profiles along the $x$- and $y$-axes.

\begin{figure*}[htbp]
	\centering
	\includegraphics[width=15cm,height=4.5cm]{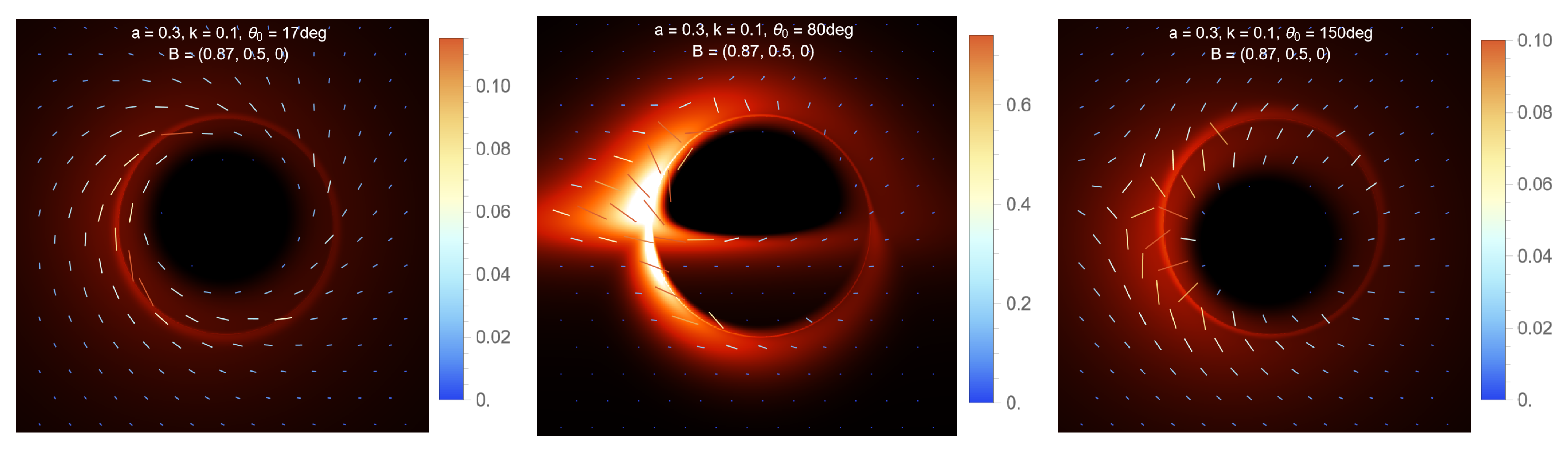}
	\includegraphics[width=15cm,height=4.5cm]{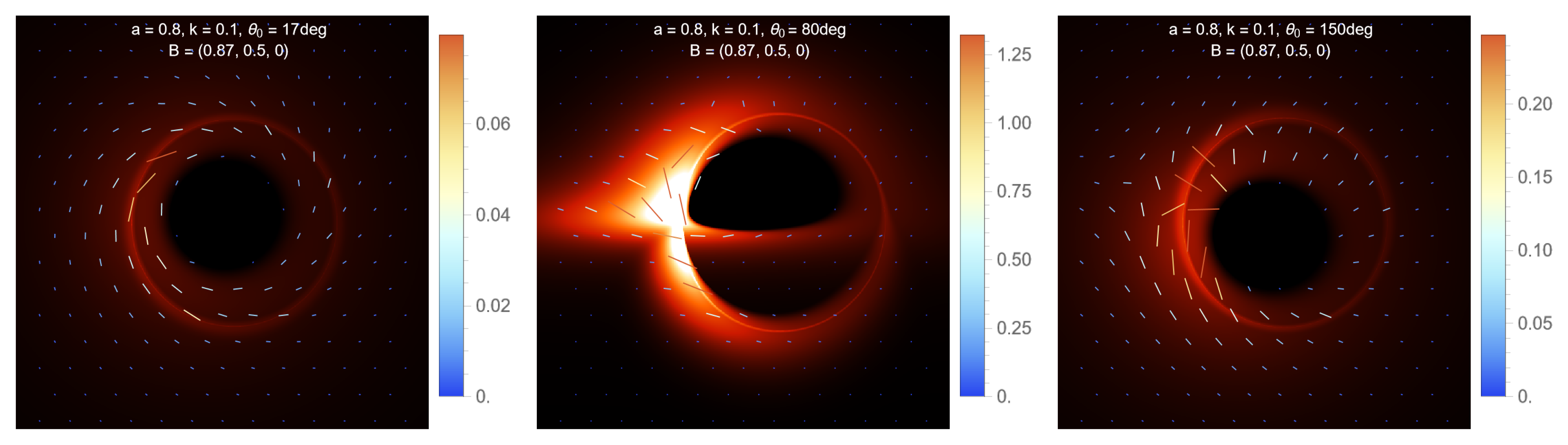}
	\includegraphics[width=15cm,height=4.5cm]{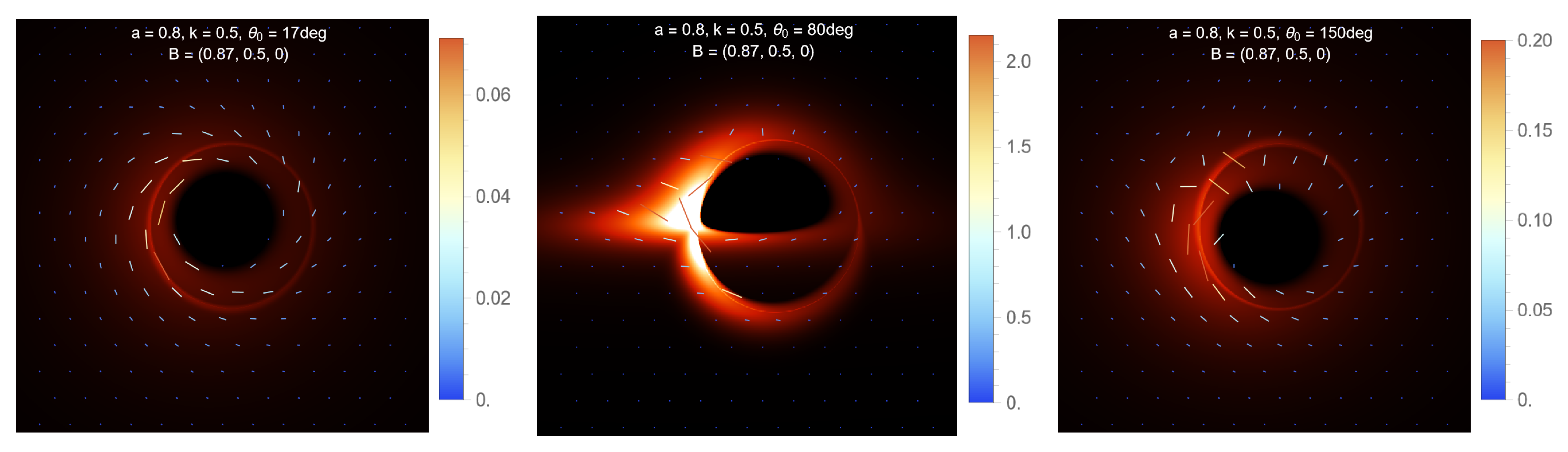}
	\caption{\label{fig:4} Polarization maps of a prograde accretion disk orbiting a BH ($M=1$) embedded in PFDM, viewed from different inclination angles $\theta_0$ (left to right: $17^\circ, 80^\circ, 150^\circ$). The maps are obtained assuming the magnetic field configuration $b2$. The rows correspond to different BH spin $a$ and PFDM intensity parameter $k$: (Top row) $a=0.3, k=0.1$; (Middle row) $a=0.8, k=0.1$; (Bottom row) $a=0.8, k=0.5$.}
\end{figure*}

Taken together, these figures reveal three robust trends driven by viewing geometry, spin, and PFDM.
\begin{enumerate}
	\item The polarized intensity exhibits a clear inclination dependence. As the viewing angle approaches an equatorial orientation ($\theta_0 \to 90^\circ$), the emission becomes increasingly asymmetric and radially extended, with the brightest polarized region concentrated on the left side of the disk—coinciding with the Doppler-brightened, approaching side of the flow. This behavior reflects the combined effect of relativistic aberration and the preferential projection of the poloidal field onto the observer's screen at high inclinations.
	
	\item At high inclination angles, increasing the BH spin $a$ systematically enhances the polarized flux near the projected photon ring. Faster rotation boosts the approaching side of the disk and intensifies light bending in the strongly rotating region, thereby magnifying both the polarized intensity and the apparent organization of the EVPA pattern in the bright-left quadrant. This spin-driven modulation primarily arises from relativistic Doppler boosting and the increased frame dragging acting on the projected magnetic field.
	
	\item PFDM leads to systematic modifications in the polarization structure by altering photon propagation in the near-horizon region. In the specific cases considered here, comparing different values of the PFDM parameter $k$ reveals clear differences in both polarized intensity and EVPA distribution across the image. These changes arise from PFDM-induced modifications to null geodesics and gravitational light bending, which in turn affect how the prescribed magnetic-field geometry is projected onto the observer’s screen. At fixed magnetic configuration, variations in $k$ therefore leave a distinct imprint on the polarization morphology through spacetime effects.
\end{enumerate}
Overall, the $b2$ configuration demonstrates that both BH spin and PFDM leave clear, distinguishable imprints on the polarization map: spin amplifies and sharpens the polarized structure near the photon ring, while PFDM redistributes polarized flux and increases EVPA contrast by altering the underlying geodesic flow.
\par
Fig.~6 extends the visualization strategy to the remaining three magnetic configurations ($b1$, $b3$, and $b4$), illustrating how the EVPA structure evolves under different viewing geometries within the specific context of a BH immersed in PFDM. Compared with the reference case $b2$, these models exhibit distinct polarization behaviors that map the interplay between their intrinsic magnetic topologies and the strong gravitational lensing characteristic of the PFDM metric. The EHT-adopted configuration $b1$ presents a strikingly discontinuous EVPA pattern. The polarization structure reveals a sharp dichotomy: in the immediate vicinity of the BH, the strong gravitational lensing and frame-dragging effects severely twist the polarization vectors associated with the inner photon ring; conversely, the emission from the outer accretion flow remains relatively insensitive to these relativistic effects. This mismatch results in a lack of global cohesion, where the highly warped inner EVPA vectors fail to align smoothly with the outer field, leading to a ``scrambled" appearance with reduced large-scale coherence. In contrast, configuration $b3$ exhibits a highly stable and ordered response. The intrinsic field lines, being aligned with the system's symmetry axis, are more robust against the twisting effects of parallel transport along the curved geodesics. Consequently, $b3$ produces a radially sweeping EVPA pattern that maintains remarkable smoothness and continuity across the image plane, even as the observer's inclination changes. Most illuminating is the behavior of configuration $b4$, which features a static dipolar field confined strictly to the accretion disk. The ordered curvature of the dipolar field lines within the emission region efficiently organizes the polarization vectors into a continuous, spiral-like structure. This result demonstrates that the emergence and continuity of spiral EVPA patterns are governed primarily by the presence of a coherent poloidal magnetic orientation within the emitting region, rather than by the detailed global extent of the magnetic field. Collectively, these comparisons underscore the decisive role of the poloidal magnetic component in shaping the observed polarization. While toroidal-dominated fields ($b1$) tend to produce stochastic or discontinuous patterns due to the differential lensing of azimuthal vectors, configurations enriched with poloidal structure ($b2, b3, b4$) successfully organize the polarization into coherent, large-scale patterns. This suggests that the organized EVPA morphologies observed in sources like M87$^*$ may be naturally associated with the presence of ordered vertical magnetic fields in the underlying accretion flow.

\begin{figure*}[htbp]
\centering
\includegraphics[width=5.5cm,height=4cm]{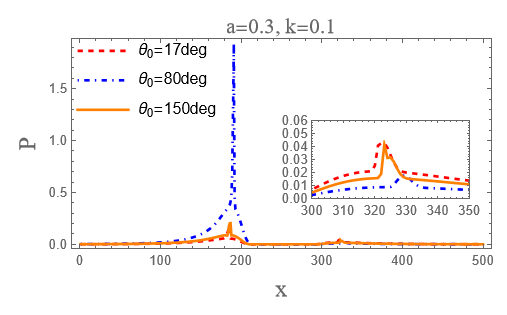}
\includegraphics[width=5.5cm,height=4cm]{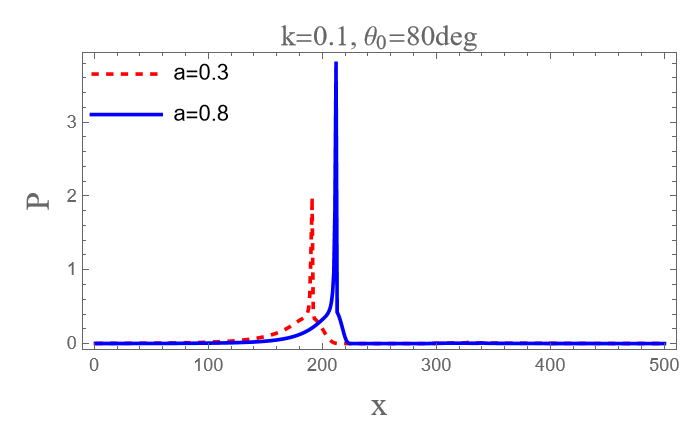}
\includegraphics[width=5.5cm,height=4cm]{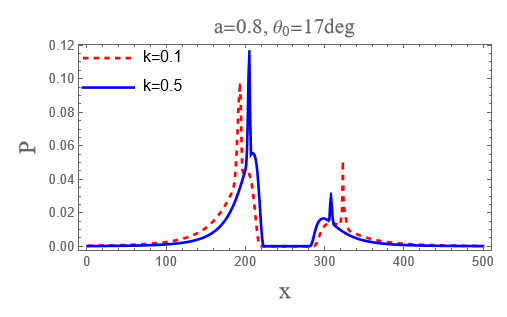}
\includegraphics[width=5.5cm,height=4cm]{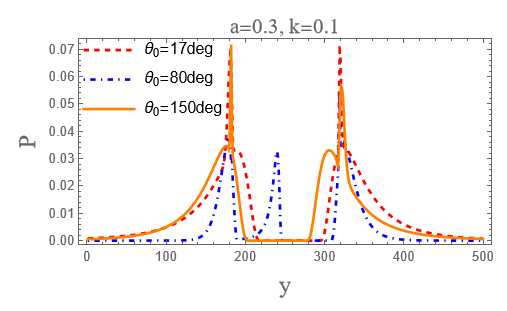}
\includegraphics[width=5.5cm,height=4cm]{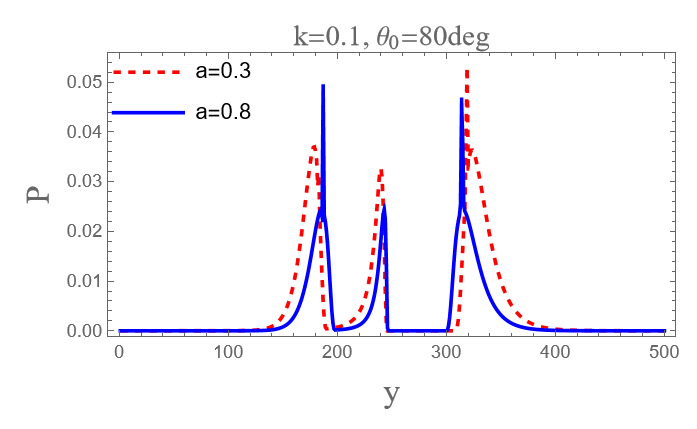}
\includegraphics[width=5.5cm,height=4cm]{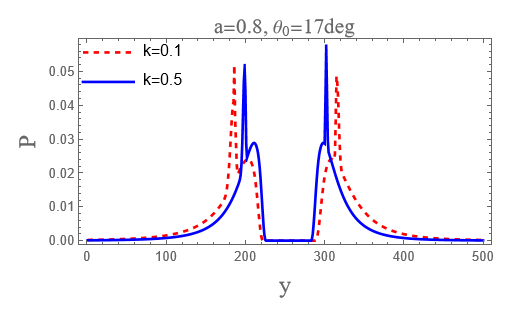}
\caption{\label{fig:5} Polarization intensity profiles along the observer's $x$-axis (Top row) and $y$-axis (Bottom row) for the magnetic field configuration $b2$. Three parameter sets are shown: (Left panel) $a=0.3, k=0.1$; (Middle panel) $k=0.1, \theta_0=80^\circ$; (Right panel) $a=0.8, \theta_0=17^\circ$.}
\end{figure*}

\begin{figure*}[htbp]
\centering
\includegraphics[width=15cm,height=4.5cm]{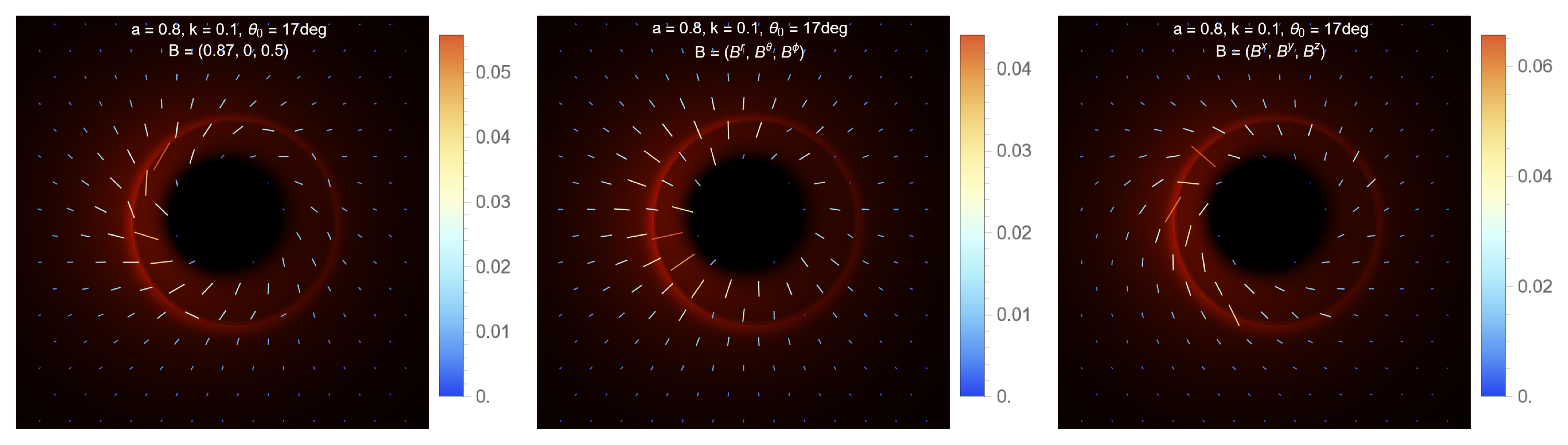}
\includegraphics[width=15cm,height=4.5cm]{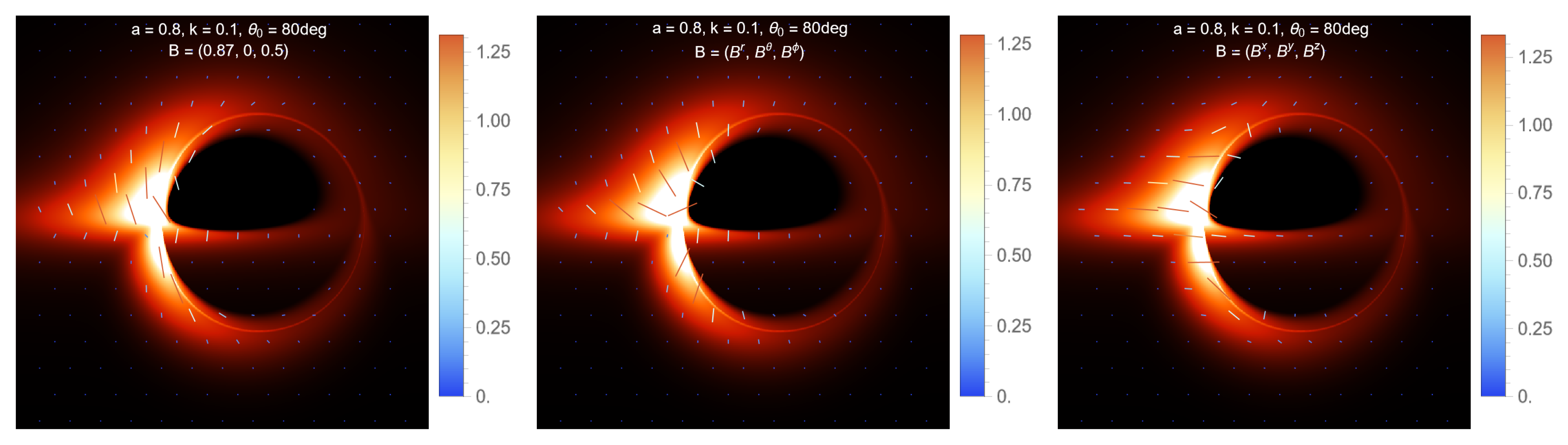}
\caption{\label{fig:6} Same as Fig.~\ref{fig:3} for $a=0.8$, $k=0.1$ and magnetic fields $b1$, $b3$, $b4$, with $\theta_0=17^\circ$ (Top row) and $\theta_0=80^\circ$ (Bottom row).}
\end{figure*}

\subsection{Role of Higher-Order Images}
\label{sec:3-3}

\begin{figure*}[htbp]
	\centering
	\includegraphics[width=15cm,height=4.5cm]{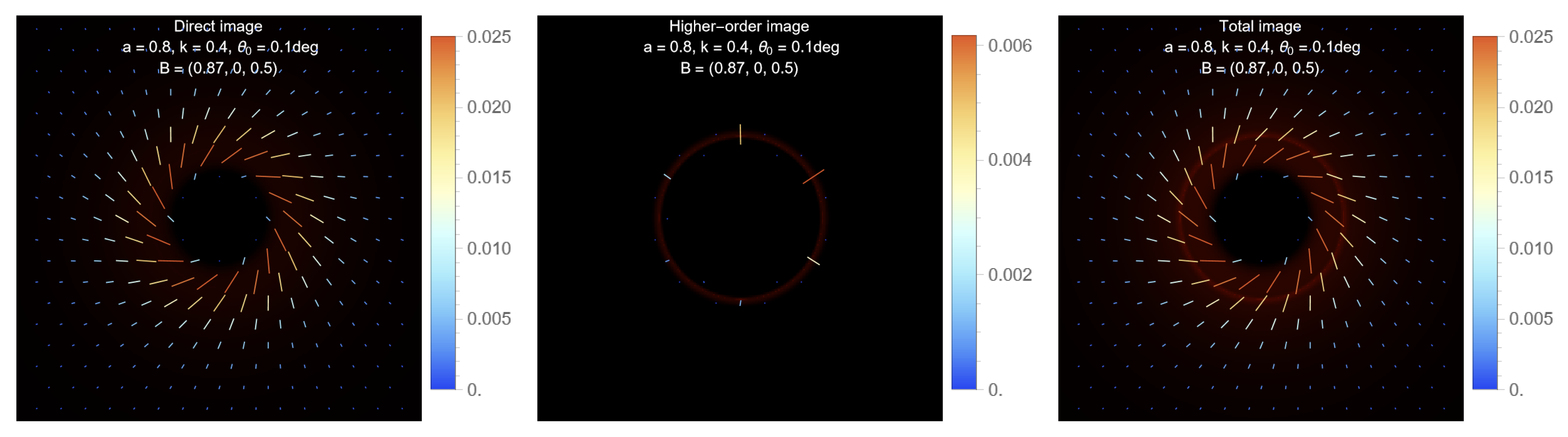}
	\includegraphics[width=15cm,height=4.5cm]{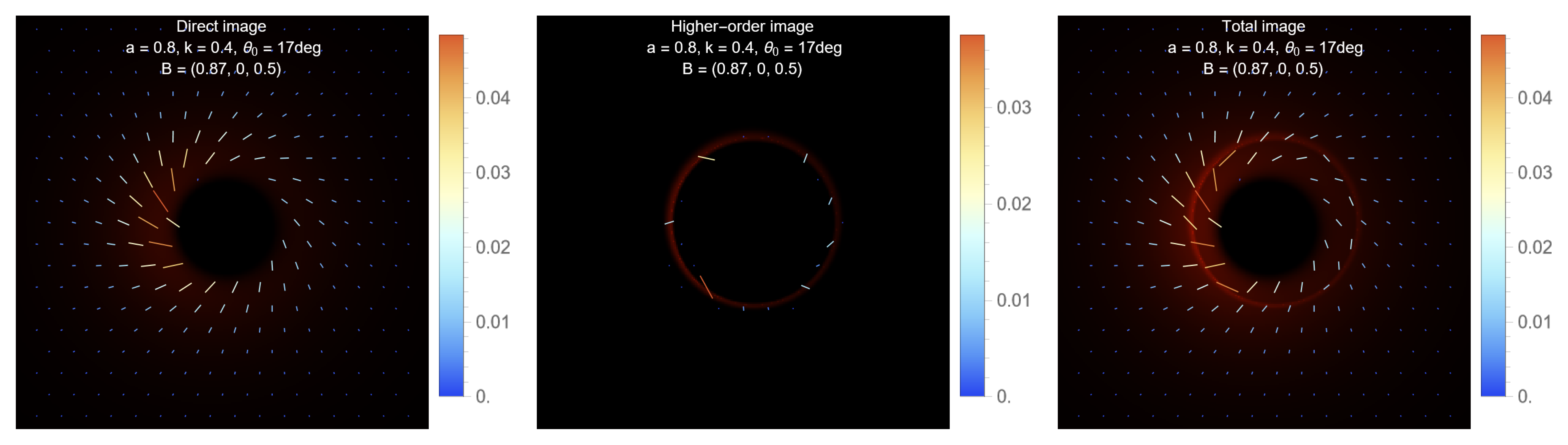}
	\caption{\label{fig:7} Decomposition of the polarization maps into the direct image, higher-order image, and total image for a prograde accretion disk orbiting a BH embedded in PFDM. The maps are obtained assuming the magnetic field configuration $b1$. The adopted parameters are $a = 0.8$ and $k = 0.4$. The rows correspond to different inclination angles: (Top row) $\theta_0=0.1^\circ$; (Bottom row) $\theta_0=17^\circ$. The columns correspond to the direct image, higher-order image, and total image, respectively.}
\end{figure*}

\begin{figure}[htbp]
	\centering
	\includegraphics[width=6.5cm,height=5cm]{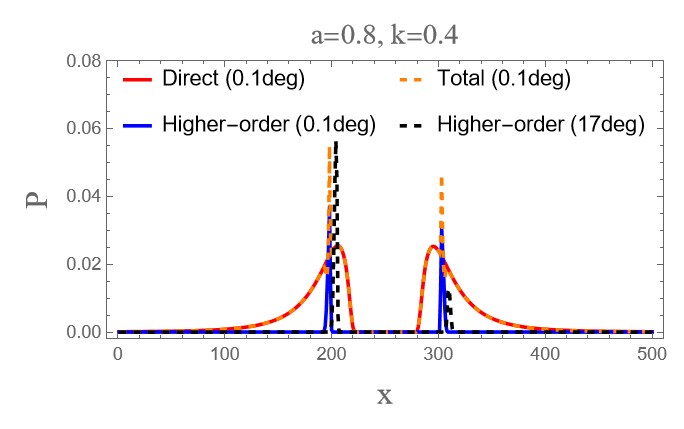}
	\caption{\label{fig:8} Polarization intensity profiles along the observer's $x$-axis for the magnetic field configuration $b1$. The adopted parameters are $a = 0.8$ and $k = 0.4$. Two inclination angles are shown: $\theta_0=0.1^\circ$ and $\theta_0=17^\circ$. For $\theta_0=0.1^\circ$, the direct, higher-order, and total contributions are plotted separately, while for $\theta_0=17^\circ$, only the higher-order contribution is shown.}
\end{figure}

Since polarization direction of a higher-order image may provide an important diagnostic for future horizon-scale observations, and may also affect the interpretation of the total polarization structure, it is useful to examine their contribution separately. Therefore, we decompose the total image into its direct and higher-order components for a representative case with \(a=0.8\), \(k=0.4\), and the \(b1\) magnetic-field configuration. The corresponding polarization maps are displayed in Fig.~7.
It shows that the total image is globally controlled by the direct component, which determines the large-scale polarization morphology across most of the image plane. The higher-order component, by contrast, is confined to a much narrower lensed region near the photon ring. Its contribution is therefore spatially localized rather than global. Nevertheless, because this higher-order signal is superposed on the direct image precisely in the strong-lensing region, it still leaves a visible imprint on the total polarization structure.

Fig.~8 shows the polarized-intensity profiles \(P\) extracted along the horizontal direction. 
The polarized-intensity profile of the direct image exhibits broad peaks associated with the main emitting region, whereas the higher-order contribution appears as much narrower features concentrated near the projected photon-ring position. The total profile is not simply identical to that of the direct image: instead, it contains localized enhancements produced by the superposition of the higher-order component on the broader direct-image background. In this sense, the higher-order image acts as a narrow correction layered on top of the dominant direct contribution. The principal consequence of this correction is not a change in the overall polarization morphology, but a local modification of its continuity. Since the polarization directions of the direct and higher-order components are generally not identical, their superposition introduces localized distortions into the total EVPA pattern. The total image therefore remains globally similar to the direct image, but no longer preserves its smooth polarization continuity in the near-ring region with the same degree of coherence.

The visibility of this effect depends strongly on the viewing inclination. For the nearly face-on case, \(\theta_0=0.1^\circ\), both the direct and higher-order polarization patterns remain relatively ordered in our uniform-field model, while the higher-order image itself is restricted to an extremely small region. As a result, once the two components are combined, the higher-order contribution is difficult to distinguish directly in the total image. Although it is already included in the total signal, its influence is largely masked by the low inclination, the ordered magnetic geometry, and the limited spatial extent over which the higher-order component appears.

At the larger inclination, \(\theta_0=17^\circ\), the situation changes qualitatively. Strong-field lensing, projection effects, and relativistic brightness asymmetry become more pronounced, which enhances the difference between the direct and higher-order components. The higher-order image becomes more asymmetric across the ring, and this asymmetry is also reflected in Fig.~8 through the increasingly uneven left--right structure of the polarized-intensity profile. When superposed on the direct image, the higher-order component therefore produces a much more evident disturbance of the polarization continuity in the total map.

Taken together, Figs.~7 and 8 show that the direct image sets the global polarization morphology, while the higher-order image provides a localized but non-negligible correction near the photon ring. At very low inclination this correction is difficult to isolate directly in the total image, whereas at larger inclination the combined effects of lensing and relativistic asymmetry make its influence significantly more apparent. These results indicate that higher-order polarized emission, although subdominant in spatial extent, can play an important role in shaping the fine polarization structure of horizon-scale images.

\section{Quantitative Polarization Signatures and Observational Context}
\label{sec:4}
The above analysis 
show that the PFDM parameter $k$ modifies the polarized image in a systematic way.
 To quantitatively assess such a modification and make comparison of the theoretical BH image to the ETH observatons of 
M87$^{*}$, it is necessary to connect the spacetime parameter not only to a characteristic dark-matter density scale, but also to the global organization of the polarization field on the image plane. In this section, we make this connection explicit. We first relate $k$ to a representative physical density scale for M87$^{*}$, then quantify the corresponding changes in the image-domain polarization observables, and finally show how these changes are realized in the resolved morphology of the polarized ring.

\subsection{Physical density scale associated with the PFDM parameter}
\label{sec:4-1}
Above analysis is in geometrized units ($G = c = 1$). For comparison of the theoretical results with the observations, it is useful to adopt the mass-density scale for a specific source. 
Since the present analysis is aimed at the horizon-scale polarization of M87$^{*}$, it is useful to rewrite the density in Eq.~(\ref{eq:7}) in cgs units by restoring the gravitational length scale $r_g=GM_{\rm BH}/c^2$ associated with the black-hole mass. Thus, we express the radial coordinate, spin parameter, and PFDM parameter in units of $r_g$, namely $r=\bar r\,r_g$, $a=\bar a\,r_g$, and $k=\bar k\,r_g$, where the barred quantities are dimensionless. Substituting these relations into Eq.~(\ref{eq:7}) yields the PFDM density in cgs units as
\begin{equation}
	\label{eq:34}
	\rho_{\rm DM}^{\rm cgs}(\bar r,\theta)
		=
		\frac{c^2}{G r_g^2}
		\frac{\bar k\,\bar r}
		{8\pi\left(\bar r^2+\bar a^2\cos^2\theta\right)^2}.
\end{equation}

\begin{table}[t]
	\caption{Representative PFDM density scale for M87$^{*}$ in cgs units, evaluated on the equatorial plane at $r=6r_g$ and $r=10r_g$.}
	\label{tab:1}
	\begin{ruledtabular}
		\begin{tabular}{ccc}
			$k$ & $\rho_{\rm DM}(6r_g)$~($\mathrm{g\,cm^{-3}}$) & $\rho_{\rm DM}(10r_g)$~($\mathrm{g\,cm^{-3}}$) \\
			\hline
			0.1 & $2.68\times10^{-7}$ & $5.79\times10^{-8}$ \\
			0.2 & $5.36\times10^{-7}$ & $1.16\times10^{-7}$ \\
			0.3 & $8.04\times10^{-7}$ & $1.74\times10^{-7}$ \\
			0.4 & $1.07\times10^{-6}$ & $2.31\times10^{-7}$ \\
			0.5 & $1.34\times10^{-6}$ & $2.90\times10^{-7}$ \\
		\end{tabular}
	\end{ruledtabular}
\end{table}

For M87$^{*}$, adopting $M_{\rm BH}=6.5\times10^9\,M_\odot$ \cite{EventHorizonTelescope:2019dse}, this expression gives the corresponding physical density scale once the location in the emitting region is specified. On the equatorial plane, $\theta=\pi/2$, Eq.~(\ref{eq:34}) reduces to
\begin{equation}
	\label{eq:35}
	\rho_{\rm DM}^{\rm cgs}(\bar r,\pi/2)
		=
		\frac{c^2}{8\pi G r_g^2}\,
		\frac{\bar k}{\bar r^3}.
\end{equation}

Equation~(\ref{eq:35}) shows that, on the equatorial plane, the PFDM density follows an $r^{-3}$ profile, while the parameter $k$ sets the overall normalization of the density scale. Because the density is position dependent, a given value of $k$ does not correspond to a unique local density. Table~I therefore lists representative values of $\rho_{\rm DM}$ at $r=6r_g$ and $r=10r_g$, two characteristic radii spanning the bright inner ring and its immediate surroundings. For the parameter range considered here, the inferred densities are substantially higher than both the dark-matter densities usually discussed on galactic scales and the plasma density typically inferred for the emitting region of M87$^{*}$. At the same time, astrophysical studies of supermassive black holes suggest that the central dark-matter distribution can be substantially enhanced relative to the large-scale halo profile through the formation of a density spike, while the actual inner profile may remain sensitive to the growth history and environment of the black hole \cite{Chan:2024yht,Shen:2023pan}. In this sense, the values listed in Table~I should be understood as representative strong-field density scales associated with the adopted PFDM parameter range, describing a physically possible regime in which dark-matter-induced modifications of the spacetime can leave observable imprints on the polarized image of M87$^{*}$.

\begin{figure*}[htbp]
	\centering
	\includegraphics[width=4.5cm,height=4.5cm]{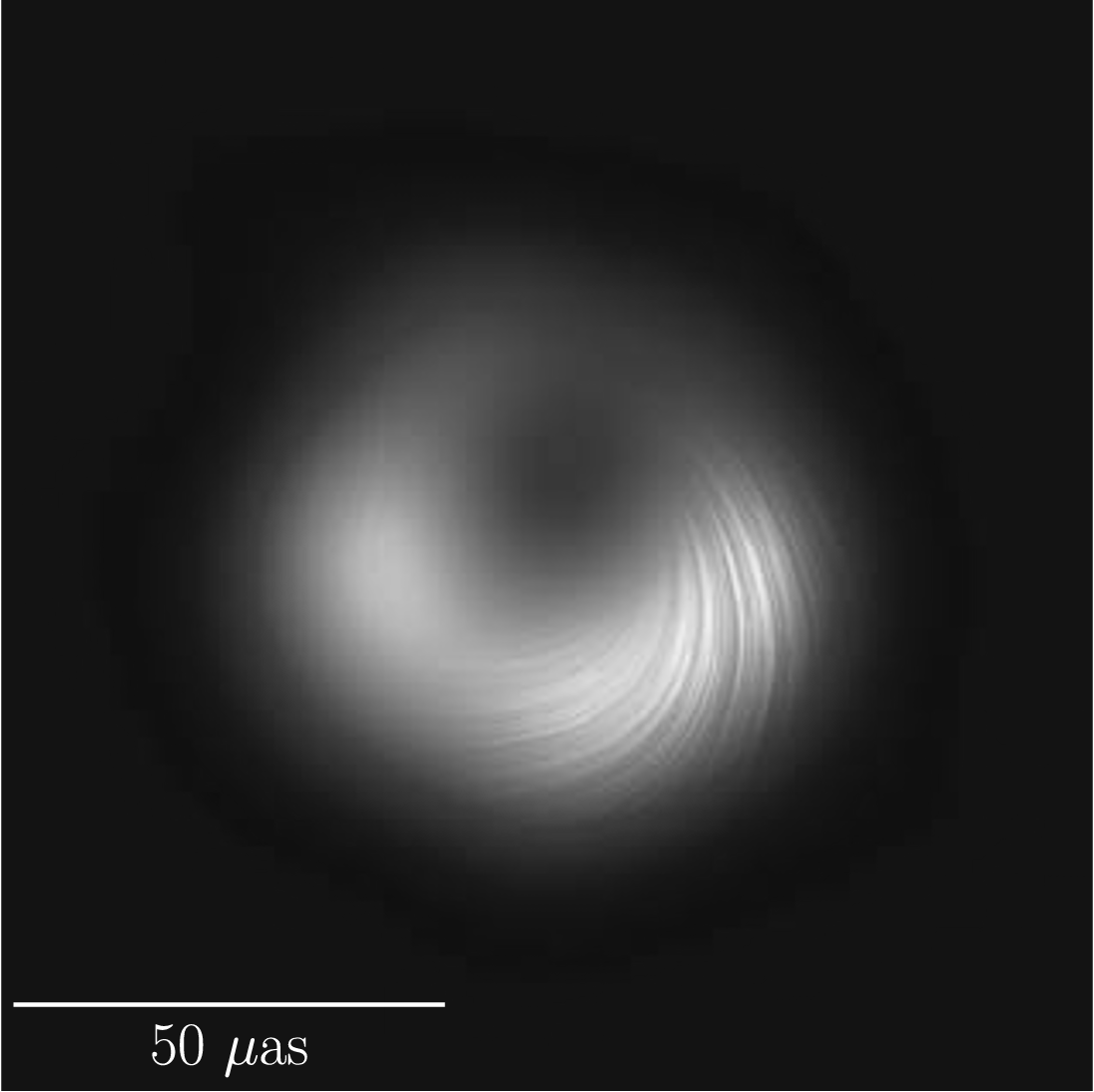}
	\includegraphics[width=4.5cm,height=4.5cm]{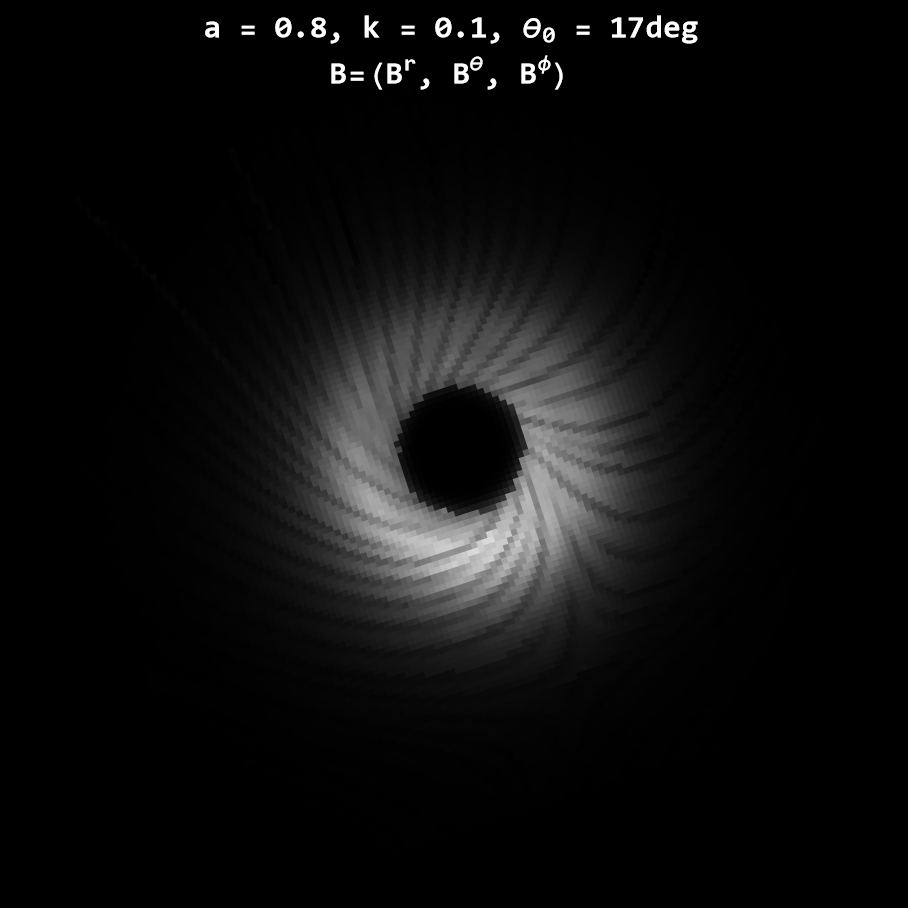}
	\includegraphics[width=4.5cm,height=4.5cm]{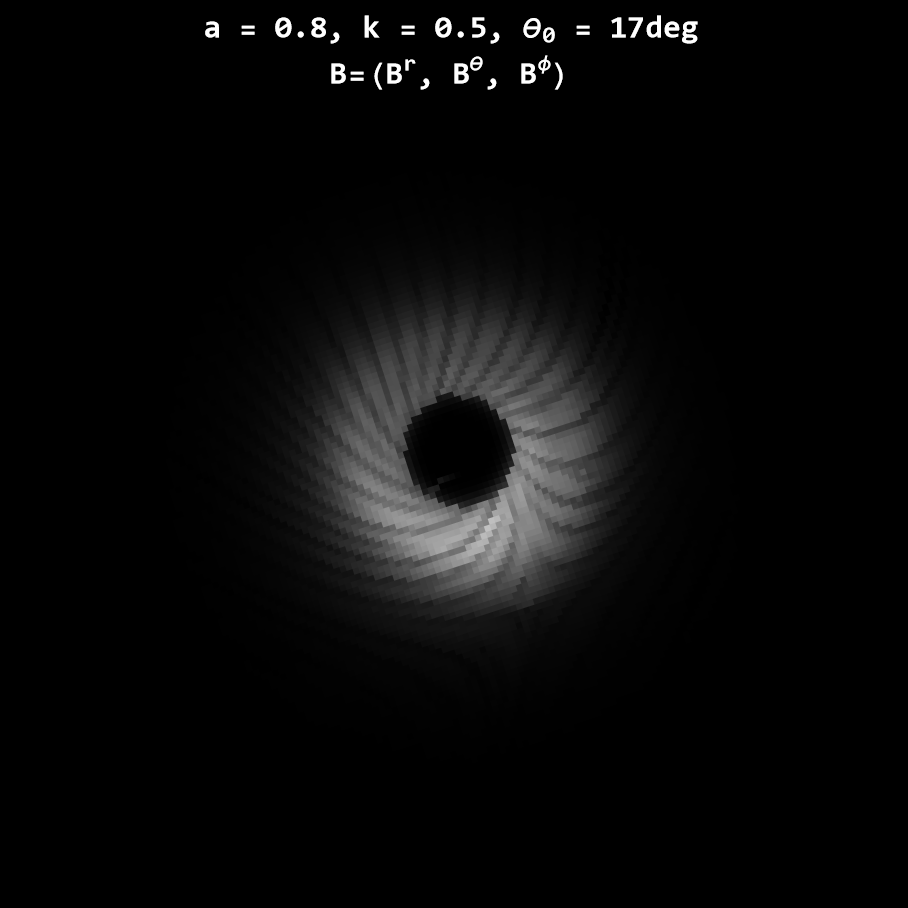}
	\includegraphics[width=4.5cm,height=4.5cm]{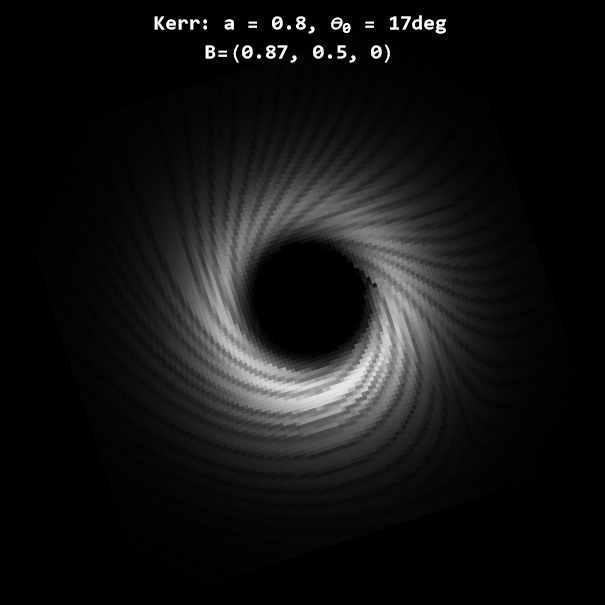}
	\includegraphics[width=4.5cm,height=4.5cm]{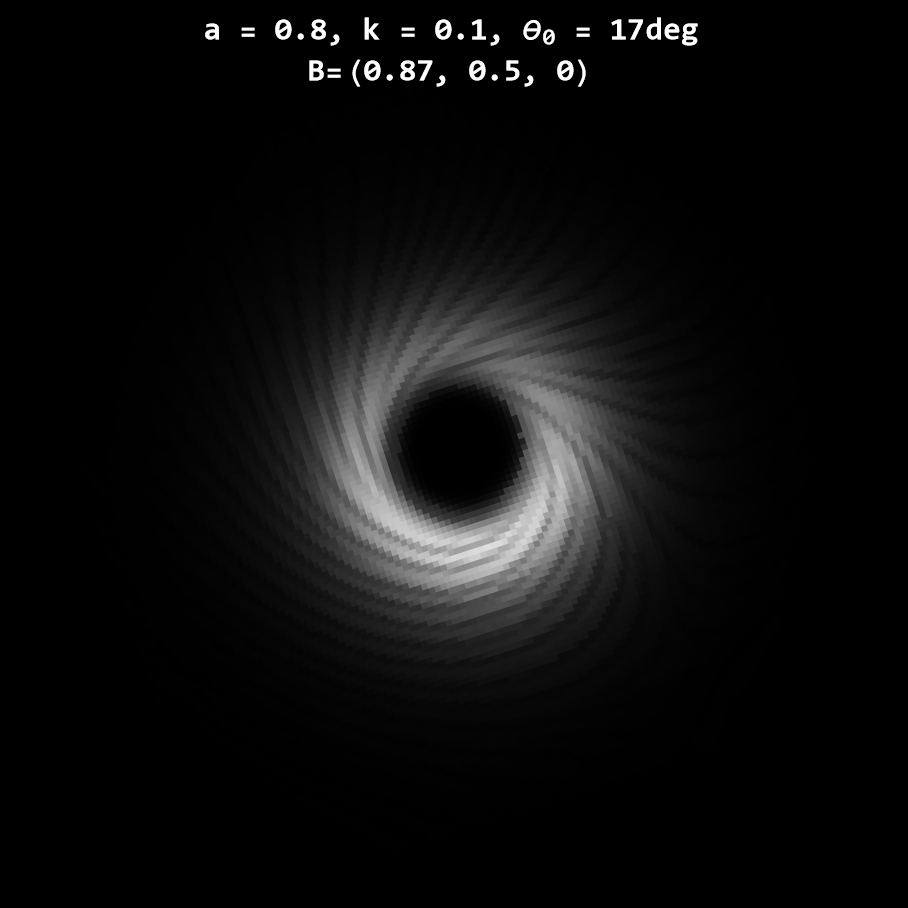}
	\includegraphics[width=4.5cm,height=4.5cm]{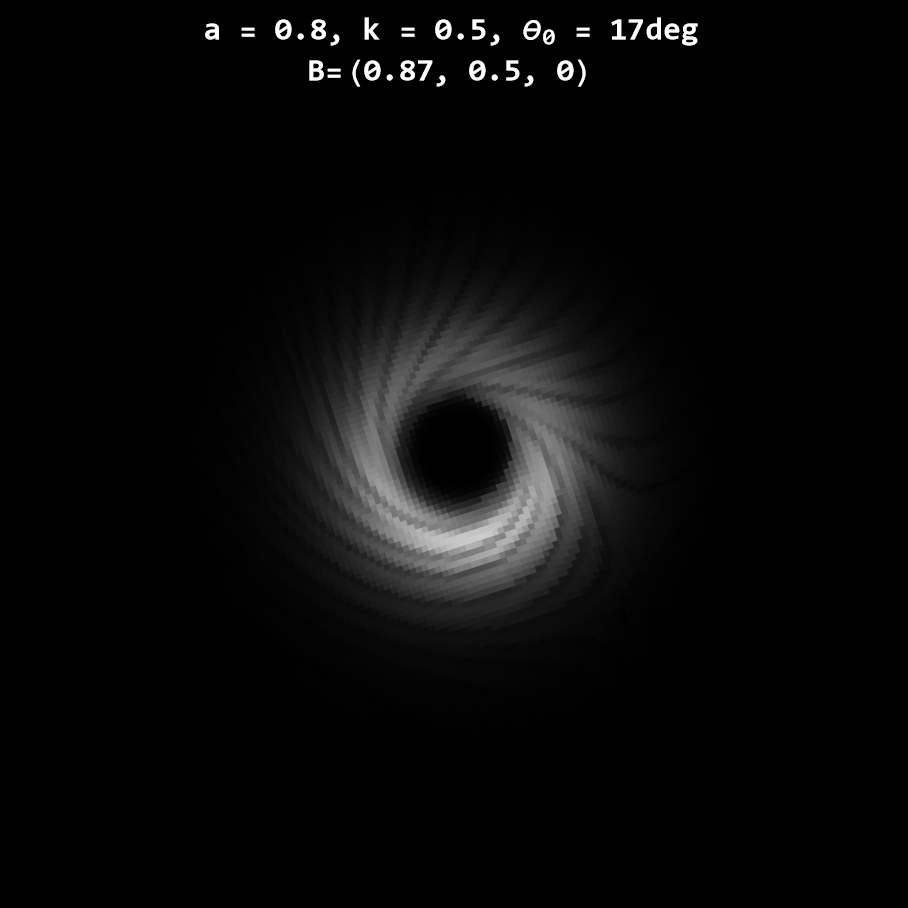}
	\caption{\label{fig:9}
		Comparison of the simulation results with the M87$^{*}$ image. Top row: (Left) the M87$^{*}$ image from Ref.~\cite{EventHorizonTelescope:2021bee}, shown here in grayscale; (Middle) the simulation result with $k=0.1$ for the magnetic-field configuration $b4$; (Right) the simulation result with $k=0.5$ for the magnetic-field configuration $b4$. Bottom row: (Left) the Kerr image; (Middle) the simulation result with $k=0.1$ for the magnetic-field configuration $b2$; (Right) the simulation result with $k=0.5$ for the magnetic-field configuration $b2$.}
\end{figure*}

\begin{figure}[htbp]
	\centering
	\includegraphics[width=6.5cm,height=5cm]{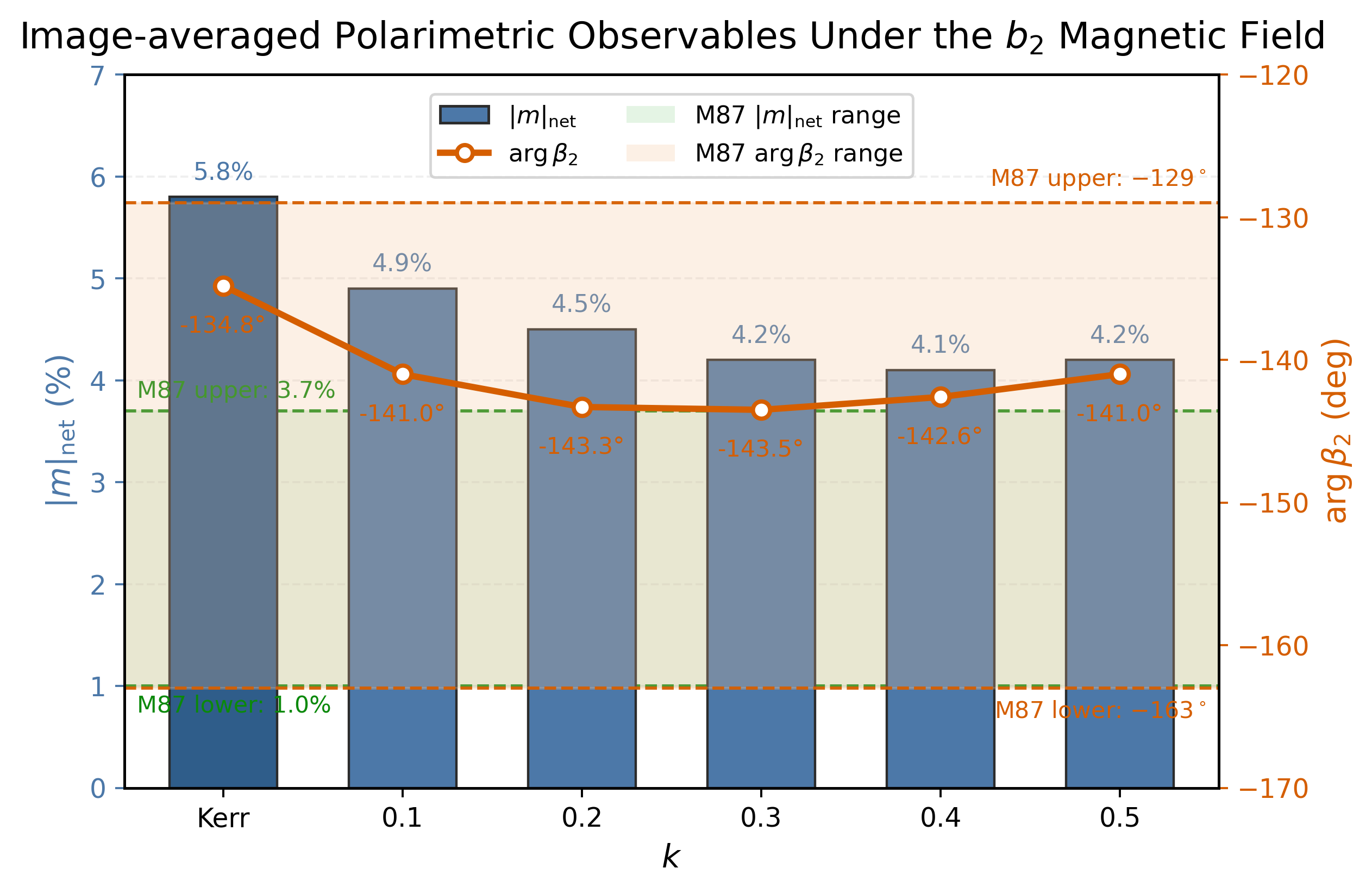}
	\caption{\label{fig:10}
		Image-domain polarimetric observables for the $b2$ magnetic-field configuration in the Kerr spacetime ($a=0.8$) and PFDM spacetimes ($a=0.8$, $k=0.1$--$0.5$). The bars show the net linear polarization fraction $|m|_{\rm net}$, while the orange curve shows the phase of the second azimuthal polarization mode, $\arg\beta_2$. The shaded horizontal bands indicate the corresponding observational ranges inferred from the EHT polarization image of M87$^{*}$, namely $1.0\%\leq |m|_{\rm net}\leq 3.7\%$ and $-163^\circ \leq \arg\beta_2 \leq -129^\circ$ \cite{EventHorizonTelescope:2021bee}.}
\end{figure}

\subsection{Image-domain polarimetric diagnostics}
\label{sec:4-2}

We now examine how the PFDM spacetime manifests itself in the polarized image. Since the EHT image of M87$^{*}$ reflects the collective organization of the polarization field over the bright ring, a meaningful comparison cannot rely on isolated local EVPA vectors alone. It is therefore useful to first inspect the large-scale morphology and then quantify its global structure through image-domain observables.

Fig.~9 shows the polarization streamline patterns for the two representative magnetic configurations, $b2$ and $b4$, after convolution with a Gaussian kernel corresponding to an effective angular resolution of $1/12$ of the field of view \cite{Gralla:2019xty}. Throughout this part of the analysis, we adopt a counterclockwise image orientation in order to facilitate comparison with the observed M87$^{*}$ polarization maps. At the morphological level, the figure shows that the gross appearance of the polarized image is controlled primarily by the magnetic-field configuration. In particular, the difference between $b2$ and $b4$ is immediately visible in the global streamline geometry: within our simplified morphological comparison, the $b2$ configuration produces a curved and more globally ordered EVPA pattern that more closely resembles the large-scale polarization organization seen in the EHT image of M87$^{*}$, especially in the bright South--West sector. By contrast, the $b4$ configuration yields a more tightly confined inner pattern. This contrast is used here only to illustrate how different magnetic topologies imprint distinct large-scale EVPA organizations in our toy model, rather than to claim a unique observational preference between the two field configurations.

By contrast, once the magnetic topology is fixed, the distinction between the Kerr and PFDM cases in Fig.~9 is visually much weaker. This is consistent with the expectation that, for a given field configuration, the large-scale polarization morphology is governed mainly by the magnetic structure itself, while the spacetime modifies the image only through its influence on light propagation and parallel transport. Consequently, the Kerr and PFDM images within the same magnetic configuration exhibit similar overall ring-like polarization patterns, and any spacetime-induced difference is too subtle to be robustly inferred from visual inspection alone. Fig.~9 thus indicates that the gross morphology is controlled primarily by magnetic topology, and also motivates the use of image-domain observables to quantify the more delicate Kerr--PFDM differences.

The first quantity we consider is the net linear polarization fraction \cite{EventHorizonTelescope:2021bee,EventHorizonTelescope:2021srq}
\begin{equation}
	\label{eq:38}
	|m|_{\rm net}
	=
	\frac{\sqrt{\left(\sum_{i}Q_{i}\right)^2+\left(\sum_{i}U_{i}\right)^2}}
	{\sum_{i}I_{i}},
\end{equation}
where the sums are over all pixels $i$ in the resolved image. Physically, $|m|_{\rm net}$ quantifies how much of the polarized signal survives global averaging, and is therefore sensitive to the cancellation of polarization vectors originating from different parts of the lensed ring.

To characterize the large-scale phase of the EVPA structure, we also consider the second azimuthal mode of the complex linear polarization \cite{EventHorizonTelescope:2021bee,EventHorizonTelescope:2021srq}
\begin{equation}
	\label{eq:39}
	\beta_2
	=
	\frac{1}{I_{\rm ann}}
	\int_{0}^{2\pi}\int_{\rho_{\rm in}}^{\rho_{\rm out}}
	\left(Q+iU\right)e^{-2i\varphi}\,\rho\,d\rho\,d\varphi,
\end{equation}
where the annular total intensity is
\begin{equation}
	\label{eq:40}
	I_{\rm ann}
	=
	\int_{0}^{2\pi}\int_{\rho_{\rm in}}^{\rho_{\rm out}}
	I(\rho,\varphi)\,\rho\,d\rho\,d\varphi .
\end{equation}
Its phase, denoted by ${\rm arg}(\beta_2) \equiv \angle\beta_2$, encodes the dominant geometric phase of the large-scale polarization field. The two quantities therefore probe complementary aspects of the image: $|m|_{\rm net}$ measures the strength of global polarization cancellation, while $\angle\beta_2$ measures the dominant phase of the image-wide EVPA morphology.

Their behavior for the $b2$ magnetic field is summarized in Fig.~10. Relative to the Kerr case, the PFDM models reduce $|m|_{\rm net}$ from $5.8\%$ to $4.1\%$--$4.9\%$. This reduction does not imply that the local polarization is uniformly suppressed in PFDM. Rather, it indicates that the polarization field is redistributed more strongly over the image plane, so that contributions from different parts of the ring cancel more efficiently after integration. In this sense, PFDM lowers the net polarization mainly by reshaping the global organization of the polarization field.

A similarly systematic effect is seen in $\arg\beta_2$. For the Kerr spacetime, we obtain $\arg\beta_2=-134.8^\circ$, whereas the PFDM cases shift this phase to approximately $-141^\circ$ to $-143.5^\circ$. Although the dependence on $k$ is not strictly monotonic across the full parameter range, the displacement relative to the Kerr baseline is persistent. This shows that PFDM changes not only the degree of global cancellation, but also the dominant phase of the resolved EVPA structure. The difference between Kerr and PFDM therefore lies less in the existence of an ordered polarized ring, which is present in both cases, than in the way this ordered pattern is globally rearranged.

The comparison with M87$^{*}$ further clarifies the role of these diagnostics. As shown in Fig.~10, the observed range of $\arg\beta_2$ ($-163^\circ \leq \arg\beta_2 \leq -129^\circ$ \cite{EventHorizonTelescope:2021bee}) encompasses both the Kerr and PFDM values obtained in our model, which means that this phase alone does not yet uniquely distinguish the two spacetimes. By contrast, the predicted values of $|m|_{\rm net}$ remain systematically above the M87$^{*}$ range ($1.0\%\leq |m|_{\rm net}\leq 3.7\%$ \cite{EventHorizonTelescope:2021bee}), but the PFDM cases shift downward relative to Kerr and therefore move in the direction favored by the data. This indicates that PFDM captures part of the large-scale polarization reorganization implied by the EHT image, although the simplified ring model adopted here does not by itself fully reproduce the low net polarization level of M87$^{*}$.

This remaining discrepancy likely points to additional depolarizing effects that are absent from the present framework, such as unresolved magnetic disorder and more complex plasma structure. These effects can all enhance polarization cancellation and may therefore be important for reducing $|m|_{\rm net}$ toward the observed range. Taken together, Figs.~9 and 10 indicate that PFDM does not generate a qualitatively new polarization morphology relative to Kerr, but instead produces a systematic shift in the global cancellation strength and the large-scale phase of the horizon-scale polarization field.

\subsection{Morphological Comparison with M87$^{*}$ and Physical Interpretation}
\label{sec:4-3}
The image-domain diagnostics discussed above show that the PFDM spacetime induces systematic changes in both the integrated polarization fraction and the large-scale phase of the EVPA field. In this subsection, we examine how these shifts are realized on the image plane and how they appear when placed alongside the EHT polarization image of M87$^{*}$. This comparison is useful because it connects the global observables to the spatial distribution of polarized emission and provides a direct geometric interpretation of the Kerr--PFDM differences. In particular, we focus on the representative case $k=0.4$, which gives one of the closest overall approaches to the observed M87$^{*}$ ranges among the PFDM models considered here.

The top row of Fig.~11 displays the fractional polarization tick maps, which characterize the azimuthal distribution of polarized emission. The EHT observation of M87$^{*}$ shows that the polarized signal is concentrated in asymmetric bright regions, particularly in the South-West sector \cite{EventHorizonTelescope:2021bee}. While both theoretical models capture this macroscopic asymmetry, the localized polarization degree remains systematically higher than observed. This is consistent with the still-elevated $|m|_{\rm net}$ values in Fig.~10. Notably, the $k=0.4$ PFDM case exhibits a localized reduction in the polarization fraction across the South-West and South arcs compared to the Kerr baseline. This suggests that the PFDM correction does not remove the ordered polarized ring, but instead redistributes the polarized signal. Physically, this redistribution arises because the modified geometry alters both the null geodesics and the parallel transport of polarization vectors, leading to a more efficient cancellation of the image-integrated Stokes $Q$ and $U$ contributions.

The bottom row explores the global EVPA topology through streamline visualizations. To ensure a consistent geometric basis, we identify the brightest simulated emission ring via a radial intensity scan and fit its ridge with a circle, which is then matched to the characteristic EHT ring diameter $d$ inferred in Ref.~\cite{EventHorizonTelescope:2019ths}. The left and right panels overlay the resulting theoretical streamlines on the processed M87$^{*}$ background \cite{EventHorizonTelescope:2021bee}.
It must be emphasized that EHT polarimetric reconstructions are intrinsically dependent on the adopted imaging pipeline and exhibit significant variability across different observational epochs \cite{EventHorizonTelescope:2021bee}. For this reason, the reconstructed background is used here only as a standardized observational reference frame, rather than as a statistical criterion for model selection. Within this common frame, the bottom row serves a limited and purely interpretive purpose: it provides a qualitative geometric visualization of how the phase information encoded in $\arg\beta_2$ is manifested in the resolved EVPA topology. In particular, the streamline comparison shows that the offset in $\arg\beta_2$ corresponds to a macroscopic azimuthal reorganization of the ordered EVPA pattern relative to the emission ring. The transition from Kerr to PFDM is therefore associated not with the emergence of a radically different polarized ring, but with a systematic reorganization of the dominant orientation of the ordered polarization field. In this sense, the streamline overlays are included only to visualize the geometric meaning of the image-domain diagnostics shown in Fig.~10.

Taken together, the analyses in this section show that the Kerr--PFDM difference is reflected consistently at three connected levels: the physical density scale associated with $k$, the image-domain polarization observables, and the resolved morphology of the polarized ring. The image-plane comparisons help visualize where the shifts in $|m|_{\rm net}$ and $\angle\beta_2$ arise geometrically, while the observables provide a compact quantitative measure of the same reorganization. In this sense, the polarimetric signature of PFDM is not the emergence of a wholly different image morphology, but a systematic redistribution of the ordered polarization field relative to Kerr. This suggests that future horizon-scale polarimetric observations may distinguish the two spacetimes most effectively by combining constraints on the integrated polarization fraction, the large-scale EVPA phase, and the azimuthal distribution of polarized emission around the ring.

\begin{figure*}[htbp]
	\centering
	\includegraphics[width=4.5cm,height=4.2cm]{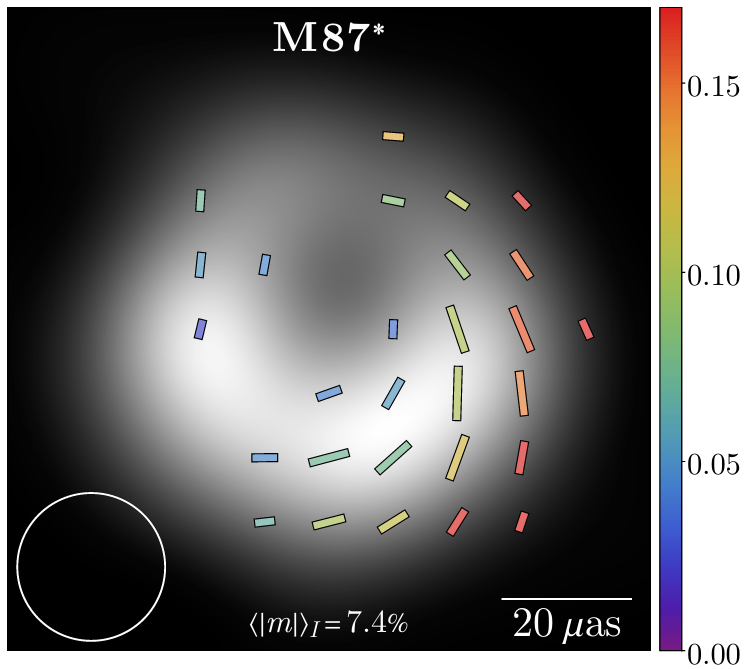}
	\includegraphics[width=4.6cm,height=4.3cm]{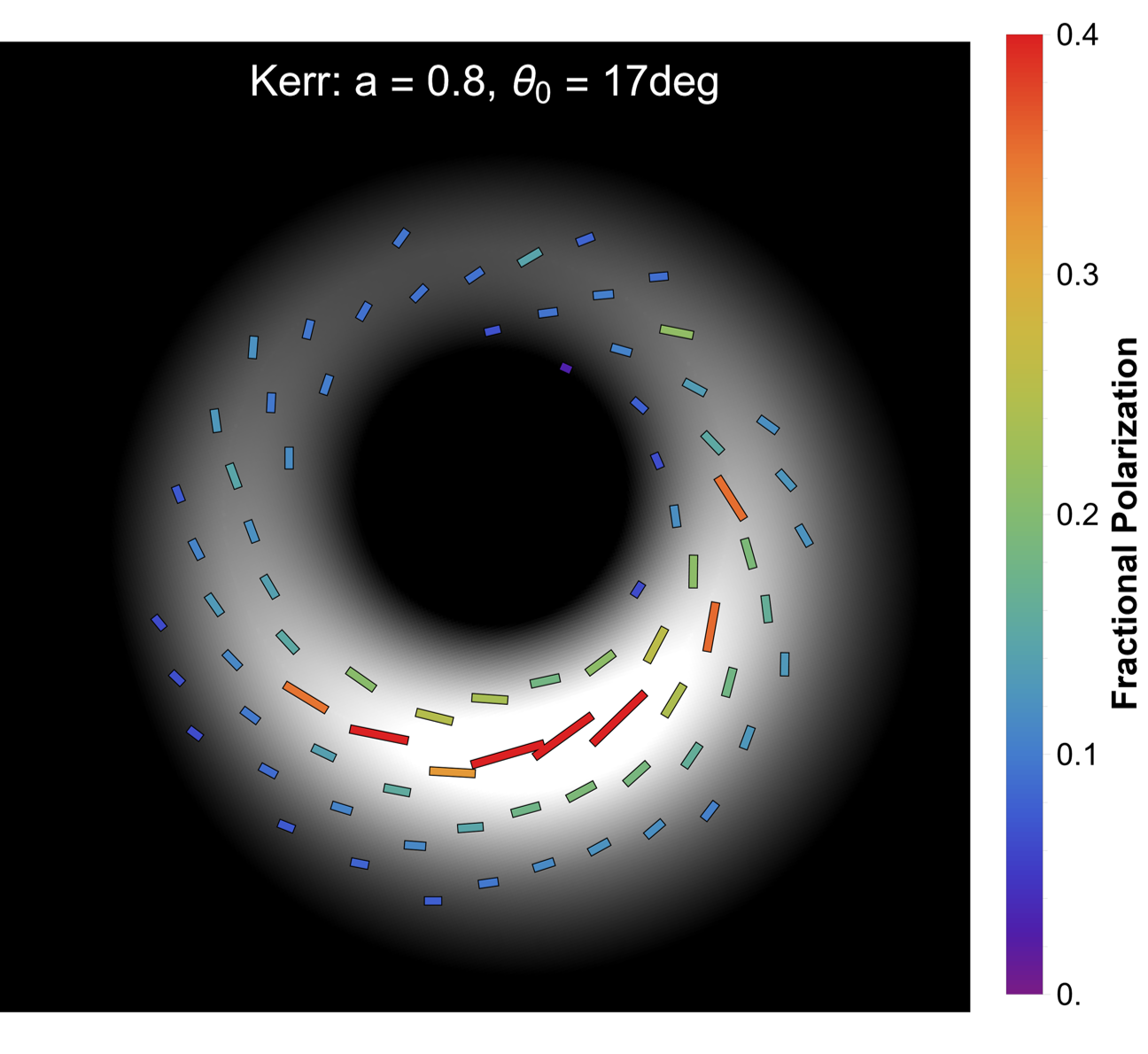}
	\includegraphics[width=4.6cm,height=4.3cm]{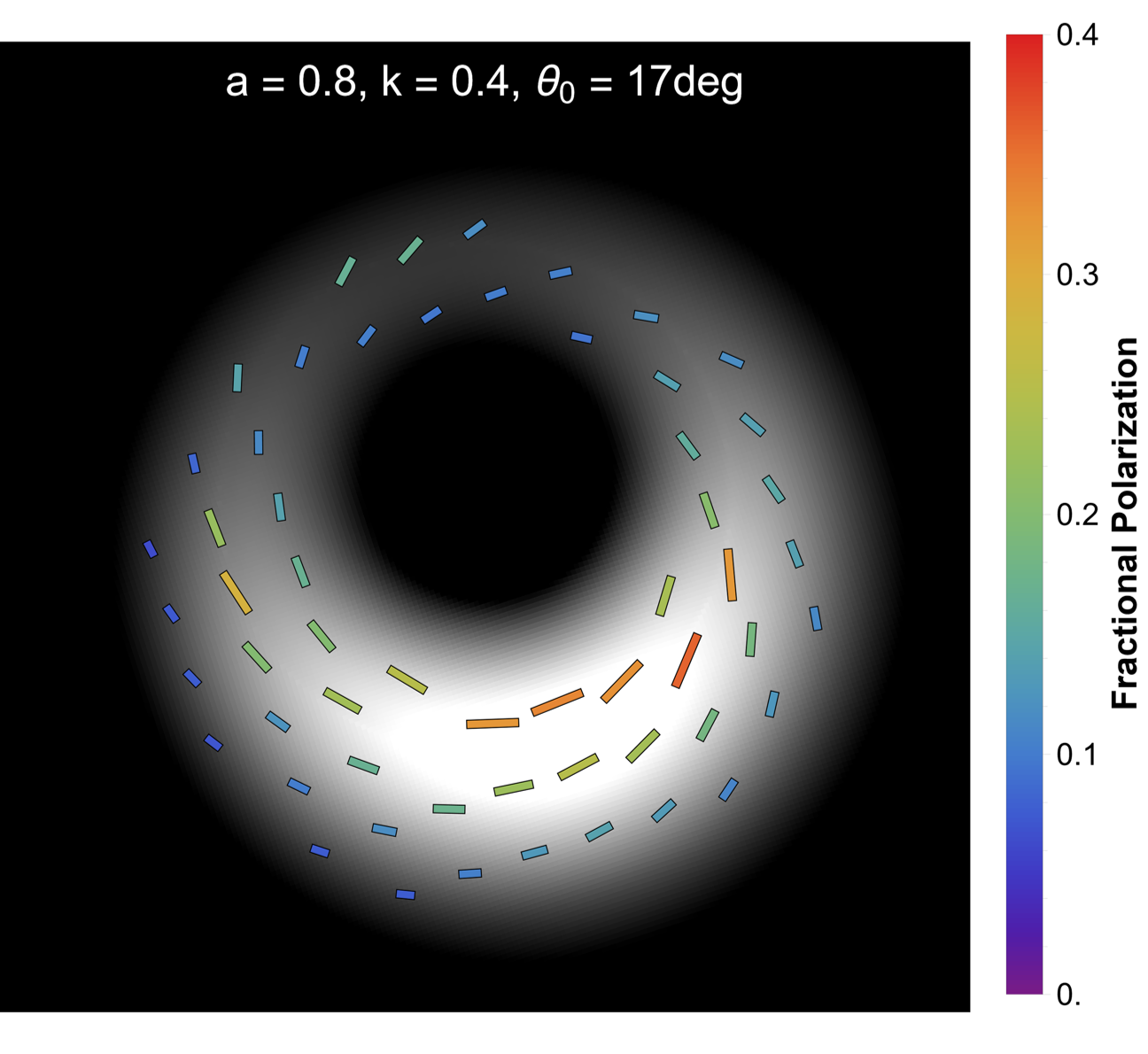}
	\includegraphics[width=4cm,height=4cm]{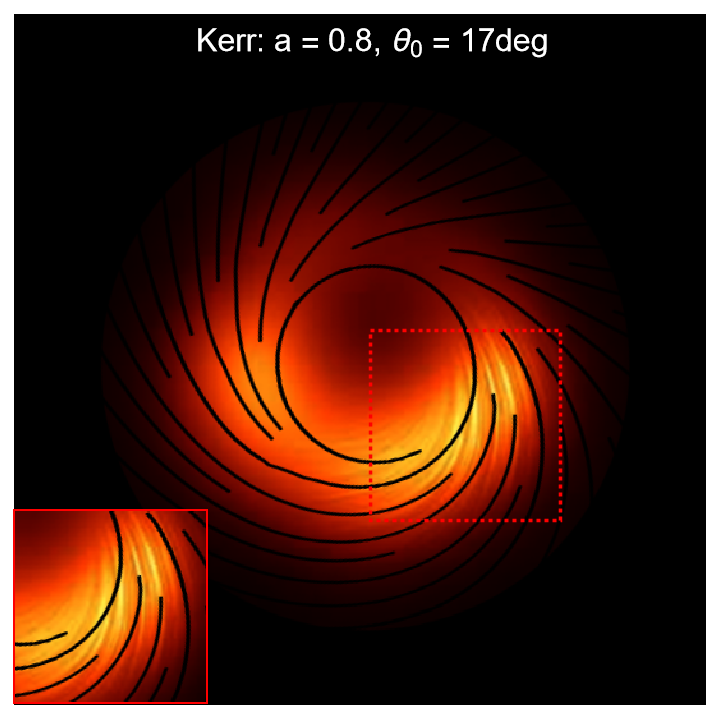}
	\includegraphics[width=4cm,height=4cm]{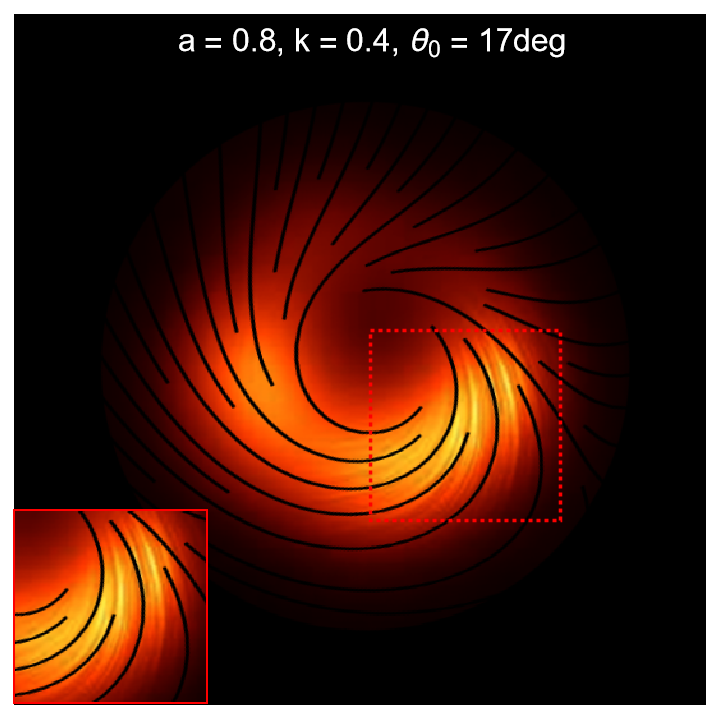}
	\caption{\label{fig:11}
		Morphological comparison between the EHT polarization image of M87$^{*}$ and representative theoretical images in the Kerr and PFDM spacetimes for the same $b2$ magnetic-field configuration. 
			Top row: (Left) Fractional-polarization tick maps of the observed M87$^{*}$ image \cite{narayan2021polarized}, (Middle) the Kerr prediction, and (Right) the PFDM prediction with $k=0.4$. The grayscale shows the total intensity, while the tick color indicates the fractional linear polarization. 
			Bottom row: Streamline visualizations. (Left) Kerr streamline pattern rendered on the same processed M87$^{*}$ background \cite{EventHorizonTelescope:2021bee}. (Right) PFDM streamline pattern with $k=0.4$ rendered on the same processed background \cite{EventHorizonTelescope:2021bee}. The top row is used for morphology-oriented comparison in the style of EHT tick plots, while the bottom row is included only to provide a qualitative geometric illustration of how the shift in $\arg\beta_2$ appears in the image plane.}
\end{figure*}

\section{Conclusions}
\label{sec:5}
\par
In this work, we have performed a systematic investigation of the polarization properties of radiation emitted from accretion flows around rotating black holes immersed in a PFDM environment. Using a fully relativistic ray-tracing framework with covariant parallel transport of the polarization vector, we explored how spacetime geometry, dark matter effects, and large-scale magnetic-field configurations jointly shape the observed polarization structure. The main purpose of this analysis is to clarify the respective roles of magnetic-field topology and PFDM-modified spacetime geometry in shaping horizon-scale polarization signatures, and to place the Kerr--PFDM comparison in a more quantitative framework.

The impact of magnetic-field geometry was examined through four representative configurations: the EHT-inspired radial--toroidal field (b1), a radial--polar configuration introduced in this work (b2), a large-scale horizon-threading poloidal field based on the analytical solution of Komissarov (b3), and a disk-anchored dipolar field (b4). Although all configurations exhibit sensitivity to spacetime and viewing parameters, the resulting polarization morphologies differ markedly. In particular, the comparison between b1 and b2 shows that replacing the azimuthal component with a polar one makes the EVPA pattern more spiral-like, whereas the stronger azimuthal contribution in b1 introduces more rapid directional changes across the image. The contrast between b3 and b4 further shows that the presence of a poloidal component alone is not sufficient to determine the polarization morphology. In b3, the more weakly varying magnetospheric field leads to limited bending of the EVPA pattern, while in b4 the stronger radial variation of the dipolar field across the emitting region produces a more pronounced spiral structure. This demonstrates that the detailed magnetic-field prescription plays a decisive role in shaping polarization patterns, beyond the influence of spacetime geometry alone.

We find that the PFDM parameter $k$ introduces substantial and systematic modifications to polarization observables. Variations in $k$ alter the near-horizon photon trajectories and the associated polarization transport, leading to pronounced changes in both the polarized intensity and the EVPA distribution. In particular, PFDM modifies the effective gravitational lensing and the apparent size of the photon ring, thereby enhancing the sensitivity of polarization morphology to strong-field spacetime curvature. The resulting EVPA patterns exhibit increased twisting and redistribution near the bright emission ring, while the overall polarization strength is modulated in a nontrivial manner. These effects are further amplified for rapidly rotating black holes and for viewing geometries close to edge-on, underscoring the coupled role of spin, inclination, and dark matter in determining the polarization structure. A decomposition of the total polarized image further shows that the direct component dominates the global polarization morphology, whereas the higher-order component provides localized but non-negligible corrections near the photon ring.

To place the PFDM parameter in a more physical context, a representative effective density scale was estimated for the M87$^{*}$ mass. This estimate should be understood only as an exploratory strong-field scale associated with the adopted PFDM parameter range, rather than as a unique astrophysical determination of the near-horizon dark-matter density. We then compared the Kerr and PFDM results with the EHT-inferred polarization characteristics of M87$^{*}$ through both morphology-oriented visualizations and the image-domain quantities $|m|_{\rm net}$ and $\arg\beta_2$. The image comparisons show that, once the magnetic topology is fixed, the distinction between Kerr and PFDM is much weaker than the contrast between different magnetic-field configurations, and is best interpreted as a geometric manifestation of a more subtle reorganization of the polarization field. This conclusion is supported quantitatively by Fig.~10: the observed range of $\arg\beta_2$ encompasses both the Kerr and PFDM values obtained in the present model, so that this phase alone does not yet uniquely distinguish the two spacetimes; by contrast, the predicted values of $|m|_{\rm net}$ remain systematically above the M87$^{*}$ range, but the PFDM cases shift downward relative to Kerr and therefore move closer to the observationally favored values. Taken together, these results indicate that PFDM does not produce a qualitatively new polarization morphology relative to Kerr, but instead introduces systematic shifts in both the global polarization cancellation strength and the large-scale phase structure of the horizon-scale polarization field.

Taken together, our results highlight the potential importance of dark matter--induced spacetime modifications in shaping horizon-scale polarization signatures. While the simplified emission and magnetic-field models adopted here are not intended to provide a definitive fit to M87$^{*}$, they show that PFDM acts as an additional strong-gravity ingredient that can systematically reorganize the ordered polarization structure on the image plane. As horizon-scale polarimetric observations continue to improve with future EHT campaigns and next-generation VLBI facilities, the framework developed in this work provides a useful avenue for jointly constraining black hole properties, magnetic-field topology, and possible dark matter effects in the innermost regions of accretion flows.

\begin{acknowledgments}
We thank the anonymous referee for helpful comments. 
This work is supported by the National Key R\&D Program (No. 2024YFA161100) , the Guangxi Science and Technology Innovation Platform Program (Leitai Action Plan, Grant No. Guike LT2600640026 ), Guangxi Key R\&D Program (Guangxi Funeng Action Plan, Grant No. FN2504240030), and National Natural Science Foundation of China (grant Nos. 12133003). E. W. L. is also supported by the Guangxi Talent Program (``Highland of Innovation Talents''). K. L. was supported by Fapesq-PB of Brazil.

\end{acknowledgments}

\nocite{*}
\bibliographystyle{unsrt}
\bibliography{apssamp}

\end{document}